\documentclass[floats,floatfix,amssymb,prd,twocolumn,superscriptaddress,nofootinbib,preprintnumbers]{revtex4-2}

\usepackage{amssymb,amsmath,verbatim,mathtools,needspace,enumitem,etoolbox,graphicx,microtype,afterpage,bm}
\usepackage[normalem]{ulem}
\usepackage[dvipsnames, usenames]{xcolor}
\usepackage{booktabs}
\usepackage[unicode, colorlinks=true]{hyperref}
\usepackage[T1]{fontenc}
\usepackage{orcidlink}
\usepackage[utf8]{inputenc}
\usepackage{siunitx}
\usepackage{placeins}

\newcommand{\sapienza}{Dipartimento di Fisica, Sapienza Università di Roma, Piazzale Aldo Moro 5, 00185, Roma, Italy}
\newcommand{\infn}{INFN, Sezione di Roma, Piazzale Aldo Moro 5, 00185, Roma, Italy}

\begin{document}

\preprint{ET-0820A-26}

\title{Impact of dynamical, quadratic, and  higher-order tidal effects\\ on gravitational-wave observations of binary neutron stars}

\author{Andrea Valletta\,\orcidlink{0009-0000-2743-8473}}
\affiliation{\sapienza}
\affiliation{\infn}

\author{Francesco Crescimbeni\,\orcidlink{0009-0001-4088-5443}}
\affiliation{\sapienza}
\affiliation{\infn}

\author{Takuya Katagiri\,\orcidlink{0000-0002-3755-3093}}
\affiliation{\sapienza}
\affiliation{\infn}

\author{Francesco Pannarale\,\orcidlink{0000-0002-7537-3210}}
\affiliation{\sapienza}
\affiliation{\infn}

\author{Paolo Pani\,\orcidlink{0000-0003-4443-1761}}
\affiliation{\sapienza}
\affiliation{\infn}

\begin{abstract}
Neutron-star binaries are prime gravitational-wave sources for current and next-generation detectors, and provide a unique probe for the behaviour of nuclear matter at supranuclear densities. Extracting information about this requires accurate waveform models that capture tidal effects during the late inspiral. We incorporate state-of-the-art tidal corrections into a post-Newtonian inspiral waveform, including dynamical tides and nonlinear contributions from quadratic Love numbers through eighth post-Newtonian (8PN) order. Using Bayesian model comparison, we find that these corrections can significantly affect the gravitational-wave signal, with the 8PN tidal response becoming comparable to lower-PN tidal contributions in the subsolar-mass regime. For exceptionally loud GW170817-like binaries observed by third-generation detectors such as the Einstein Telescope, higher-order tidal effects can be distinguished with high statistical significance, while neglecting them can introduce substantial biases in the inferred tidal deformabilities and, consequently, in constraints on the neutron-star equation of state. These corrections are particularly relevant for putative subsolar-mass binaries: the lower compactness of their constituents enhances tidal effects and makes them potentially important even for current detectors. Our results highlight the importance of higher-order tidal corrections for precision gravitational-wave measurements of dense nuclear matter.
\end{abstract}

\maketitle

\section{Introduction}

The observation of gravitational waves (GWs) from the binary
neutron star (BNS) merger GW170817 opened a new avenue for
probing the properties of matter at supranuclear densities
~\cite{LIGOScientific:2018cki}.
Unlike binary black holes, neutron stars carry information
about their internal structure through tidal interactions,
which leave characteristic imprints on the GW signal during
the inspiral. These effects provide a powerful probe of the
neutron-star equation of state (EoS), complementing
electromagnetic observations and nuclear experiments
~\cite{Flanagan:2007ix,Hinderer:2007mb,Chatziioannou:2020pqz}.

The tidal signatures observed in GW170817 provided the first
constraints on the mass-weighted tidal deformability
$\tilde{\Lambda}$, probing neutron stars with component masses
in the range $\sim 1.2-1.6\,M_\odot$ and a total mass
$M_{\rm tot}\simeq 2.74\,M_\odot$
~\cite{LIGOScientific:2018cki}.
While current measurements remain dominated by statistical
uncertainties, the increasing sensitivity of existing
detectors and the advent of third-generation (3G)
observatories, such as the Einstein Telescope (ET) and
Cosmic Explorer (CE)
~\cite{Punturo:2010zz,Maggiore:2019uih,ET:2025xjr,Reitze:2019iox},
will enable significantly more precise measurements of
neutron-star tidal properties.
In particular, the order-of-magnitude improvement in strain
sensitivity targeted by 3G detectors will yield a population
of exceptionally loud nearby BNS events, with
signal-to-noise ratios (SNRs) potentially reaching several
hundreds or more
~\cite{ET:2025xjr,Dupletsa:2026jzs}.
Although relatively rare, these observations will provide
unprecedented opportunities to constrain the EoS~\cite{Pacilio:2021jmq,Koehn:2026nin,Wouters:2026nbh} and probe
subleading tidal effects.
At the same time, the increased statistical precision will
make waveform systematics an increasingly important
limitation, requiring a more accurate description of the
binary dynamics and neutron-star finite-size effects.

Tidal interactions are commonly described within the
post-Newtonian (PN) framework, which provides a systematic
expansion of the binary dynamics in powers of the orbital
velocity. The leading tidal contribution to the GW phase
enters at fifth PN (5PN) order and is determined by the
static, electric quadrupolar tidal deformabilities of the
two stars, commonly parametrized by the mass-weighted
combination $\tilde{\Lambda}$
~\cite{Flanagan:2007ix,Vines:2011ud}.
Subleading tidal corrections arise at higher PN orders and
encode additional properties of the stellar response
(see Ref.~\cite{Chakraborty:2026qru} for a recent review).
These include higher-order corrections to the quadrupolar
interaction, higher-multipole tidal effects, tidal tails,
and spin-tidal couplings
~\cite{Vines:2011ud,Damour:2012yf,
JimenezForteza:2018rwr,Henry:2020ski,
Abdelsalhin:2018reg}.
Previous studies have shown that neglecting such
contributions can introduce systematic errors in the
inferred tidal parameters, potentially limiting the
accuracy of EoS measurements
~\cite{JimenezForteza:2018rwr,Henry:2020ski,Gupta:2023oyy}.
Assessing the importance of these corrections is therefore
essential for precision GW astronomy, particularly in the
high-SNR regime accessible to future detectors.

Recent theoretical developments have extended the
description of tidal interactions beyond the standard
linear, adiabatic approximation.
One important class of corrections arises from the
nonlinear response of neutron stars to external tidal
fields. Whereas the conventional description assumes that
the induced multipole moments depend linearly on the
applied tidal field, nonlinear interactions generate
additional contributions characterized by higher-order
response coefficients. Among them are the coefficients known as quadratic tidal Love
numbers which describe the response at
second order in the tidal field and enter the conservative
binary dynamics and GW phase at 8PN order
~\cite{Pani:2025qxs}.

A distinct but equally important contribution arises from
the dynamical tidal response.
In the adiabatic approximation, the induced multipole
moments are assumed to follow the external tidal field
instantaneously, with their amplitudes determined by static
tidal Love numbers. Beyond this approximation, the finite
response time of the stellar fluid introduces
frequency-dependent corrections, which can be systematically
described through an expansion in time derivatives of the
tidal field.
This correction also contributes to the GW
phase at 8PN order and is characterized by a dynamical
tidal Love number. Remarkably, this response includes both
an EoS-dependent finite contribution and a universal
logarithmic running term
~\cite{Apostolidis:2026qsg}.

Although formally suppressed by higher powers of the orbital
velocity, these contributions can be enhanced by the strong
dependence of tidal deformabilities on stellar compactness.
For instance, the dimensionless electric tidal deformability
of multipolar order $\ell$ scales schematically as
$\Lambda_\ell\propto k_\ell C^{-(2\ell+1)}$,
where $k_\ell$ is the corresponding tidal Love number and
$C$ is the stellar compactness.
Higher-multipole and nonlinear tidal effects can therefore
become relatively more important for less compact stars,
potentially compensating for part of their PN suppression. The same applies to quadratic~\cite{Pani:2025qxs} and dynamical~\cite{Apostolidis:2026qsg} tidal effects.
This suggests that the relevance of higher-order tidal corrections depends not only on the SNR, but also sensitively on the masses and compactnesses of the binary components.

A particularly interesting regime is represented by
subsolar-mass neutron stars, which may form through
alternative astrophysical channels, such as the
fragmentation of gravitationally unstable, neutron-rich
accretion disks around black holes formed in collapsars
~\cite{Metzger:2024ujc,Baibhav:2026iny}.
Although such objects are not expected from standard
stellar-evolution channels, dedicated GW searches have
investigated the possible existence of compact binaries
containing subsolar-mass components.
Their detection would have important implications for
compact-object formation and fundamental physics,
potentially revealing alternative formation scenarios or
exotic compact objects
~\cite{Crescimbeni:2024cwh}.
From the perspective of tidal physics, low-mass neutron
stars are especially promising because their lower
compactness leads to much larger tidal deformabilities,
potentially allowing them to be distinguished from other
subsolar-mass compact-object candidates
~\cite{Crescimbeni:2024cwh,Crescimbeni:2024qrq}.
Consequently, higher-order tidal contributions may become
significant at considerably lower SNRs than in
canonical-mass BNS systems and could be relevant even for
current-generation detectors.

Despite these theoretical and observational developments,
the measurability of the complete higher-order tidal sector,
and especially the recently derived 8PN contributions,
remains largely unexplored.
In particular, it is important to establish whether these
effects can be distinguished from lower-order tidal
interactions, how their measurability depends on the binary
parameters, and to what extent neglecting them biases the
reconstruction of neutron-star properties.
Since different tidal contributions can be partially
degenerate, their relative importance in the GW phase
does not necessarily translate directly into their
observational measurability.

In this work, we address these questions by constructing
a frequency-domain \texttt{TaylorF2} inspiral waveform
including tidal corrections through 8PN order.
In addition to the leading and subleading quadrupolar
terms, our model incorporates the 6.5PN tidal-tail and
spin-tidal contributions, as well as the leading magnetic tidal terms, the 7PN quadrupolar and
electric-octupolar corrections, and the recently derived
8PN dynamical and quadratic tidal effects
~\cite{JimenezForteza:2018rwr,Henry:2020ski,
Abdelsalhin:2018reg,Pani:2025qxs,
Apostolidis:2026qsg}.
We investigate both canonical-mass and subsolar-mass
binaries, adopting a physically motivated termination
of the inspiral waveform to avoid extrapolating the PN
description into the strongly nonlinear merger regime.

We perform Bayesian model comparisons to quantify the
measurability of the individual higher-order tidal
contributions as functions of the binary masses and SNR.
We first assess the impact of the complete 7PN sector,
including the leading electric-octupolar contribution,
and subsequently investigate the combined dynamical
and quadratic tidal corrections at 8PN order.
We also perform an agnostic analysis in which the
effective 8PN tidal coefficient is treated as an independent
parameter, rather than being fixed through
EoS-insensitive quasi-universal relations~(QURs)~\cite{Yagi:2016bkt}.
This allows us to assess its independent measurability
and examine its degeneracies with the leading tidal
parameters.
Finally, we investigate the resulting systematic errors
in the reconstruction of the individual stellar masses
and tidal deformabilities, with particular emphasis on
their implications for EoS inference.

Our results show that higher-order tidal effects can become
measurable at comparatively moderate SNRs for subsolar-mass
binaries, whereas exceptionally loud events are generally
required in the canonical neutron-star mass range.
We also find that neglecting the 8PN tidal sector can lead
to appreciable biases in the inferred tidal parameters,
even when the corresponding corrections are relatively
small compared to the leading tidal contribution.
These findings highlight the importance of incorporating
higher-order tidal physics into waveform models for
precision measurements of neutron-star matter.

The paper is organized as follows.
In Sec.~\ref{sec:waveform}, we describe the waveform model
and the tidal corrections included through 8PN order.
In Sec.~\ref{sec:results}, we present the Bayesian
framework, investigate the measurability of the higher-order
tidal contributions, and assess their impact on
neutron-star parameter estimation.
Finally, in Sec.~\ref{sec:conclusions}, we summarize our
results and discuss their implications for future
GW observations.

\section{Waveform modelling}
\label{sec:waveform}
In this section, we present the frequency-domain PN waveform model adopted in our analysis, discuss the reference frequencies defining its range of applicability, and assess the relative importance of tidal contributions at different PN orders. Unless otherwise stated, we use geometrized units $G=c=1$ throughout.

\subsection{\texttt{TaylorTidal8PN}  waveform}
\label{subsec:TF2_wf}

We consider the GW signal from a quasi-circular BNS. In the frequency domain, we model the signal using  
the standard \texttt{TaylorF2} waveform~\cite{Damour:2000zb}, augmented by tidal contributions to the phase up to $8$PN order~\cite{Wade:2014vqa, Pani:2025qxs, Apostolidis:2026qsg, Chakraborty:2026qru}.
Neglecting amplitude corrections, the gravitational waveform takes the form
\begin{equation}
\label{eq:taylorF2}
\tilde{h}(f)=A\, f^{-7/6}\, \exp\left[i\psi(f)\right],
\end{equation}
where $f$ is the GW frequency, $A\propto\left[(1+z){M}_{c}\right]^{5/6}/d_L$ is the signal amplitude, ${M}_{c}= (m_1 m_2)^{3/5}/(m_1+m_2)^{1/5}$ is the chirp mass, and $d_L$ is the luminosity distance between the GW detector and the binary. The masses entering the frequency-domain phase are detector-frame masses, which are related to the source-frame masses by $m_i^{\rm det}=(1+z)m_i$, with $i=1,2$. We define
\begin{equation}
     M=m_1+m_2,   \qquad
    \eta=\frac{m_1m_2}{M^2},
    \qquad
    \delta_{\eta}=\sqrt{1-4\eta},
    \label{eq:masses}
\end{equation}
as the total mass, the symmetric mass ratio, and mass-difference parameter, respectively. We further define the mass ratio as
\begin{align}
\label{eq:massratioq}
q={m_2}/{m_1}\leq1,
\end{align}
and the component mass fractions as
\begin{align}
    X_i=\frac{m_i}{M}.
\end{align}

The phase~$\psi$ in Eq.~\eqref{eq:taylorF2} is expanded in terms of the PN parameter~$x=(\pi f^{\rm src} M)^{2/3}$, such that a term proportional to $x^n$ in the PN expansion corresponds to the $n$PN order relative to the leading Newtonian point-particle contribution. We decompose the phase~$\psi$ into the point-particle contribution~$\psi_{\rm pp}$ and the tidal contribution~$\delta\psi_{\rm tidal}$:
\begin{align}
    \psi(x)=\psi_{\rm pp}(x)+\delta \psi_{\rm tidal}(x).
\end{align}
For the point-particle phase~$\psi_{\rm pp}$, we consider the standard nonspinning circular-orbit contribution up to $3.5$PN~\cite{PhysRevD.44.3819,Damour:2000gg,PhysRevD.71.084008,Buonanno:2009zt}, 
together with spin effects up to 4PN order, assuming that spins are aligned with the orbital angular momentum~(see Refs.~\cite{Franciolini:2021xbq,Favata:2021vhw} and references therein for details; see also, e.g., Refs.~\cite{Kidder:1992fr,PhysRevD.48.1860,Kidder:1995zr,Poisson:1997ha,Mikoczi:2005dn,Gergely:1999pd,PhysRevD.74.104034,Mishra:2016whh}). 

Spin-induced quadrupole moments are included through the approximate EoS-insensitive Q–Love relation. Defining the dimensionless spin-induced quadrupole $\kappa_i$ through 
\begin{equation}
    Q_i^{\rm spin}=-\kappa_im_i^3\chi_i^2,
    \label{eq:spin-induced}
\end{equation}
we adopt the fitting relation of Ref.~\cite{Yagi:2016bkt},
\begin{equation}
\begin{aligned}
\ln \kappa_i
=&\; 0.194 + 9.36\times 10^{-2}\ln\Lambda_2^{(i)} + 4.74\times 10^{-2}\left(\ln\Lambda_2^{(i)}\right)^2\\
&-4.21\times 10^{-3}\left(\ln\Lambda_2^{(i)}\right)^3 + 1.23\times 10^{-4}\left(\ln\Lambda_2^{(i)}\right)^4 .
\end{aligned}
\label{eq:loveQ}
\end{equation}

For the tidal contribution~$\delta \psi_{\rm tidal}$, we consider terms up to $8$PN-order term. Specifically, we include the leading-order static electric quadrupolar tidal term at $5$PN order; its next-to-leading-order electric correction and the leading-order static magnetic quadrupolar contribution at $6$PN order; the tidal-tail term and the spin-tidal couplings to the electric and magnetic quadrupolar deformabilities at $6.5$PN order; the next-to-next-to-leading-order electric quadrupolar correction and the leading-order static electric octupolar tidal term at $7$PN order; and the dynamical and quadratic tidal contributions at $8$PN order. The tidal phase can then be written schematically as
\begin{equation}
\begin{aligned}
\delta\psi_{\rm tidal}
=&
\frac{3}{128\eta x^{5/2}}
\left[
\mathcal{A}_{5}x^5
+
\mathcal{A}_{6}x^6
\right.\\
&\left.+
\mathcal{A}_{13/2}x^{13/2}+\mathcal{A}_{7}x^7
+
\mathcal{A}_{8}(x)x^8
\right],
\label{eq:tidal_full_8PN}
\end{aligned}
\end{equation}
where ${\cal A}_n$ denotes the PN coefficients at $n$PN order. Note that ${\cal A}_8$ depends on the PN parameter~$x$ through the logarithmic running term~\cite{Apostolidis:2026qsg}. In what follows, we provide the explicit expressions for ${\cal A}_n$. 

A Python implementation of the waveform approximant including all these terms --- \texttt{TaylorTidal8PN} --- is publicly available at~\cite{Valletta:TaylorTidal8PN}.

\subsubsection{$5$PN term}
The $5$PN tidal contribution contains only the static electric quadrupolar tidal term. For the star labeled by $i$, we define the corresponding dimensionless tidal deformability~$\Lambda_2^{(i)}$ through the linear relation between the induced mass quadrupole moment~$Q_{ab}^{(i)}$ and the quadrupolar electric tidal field~${\cal E}_{ab}$~\cite{Hinderer:2007mb,Flanagan:2007ix}
\begin{align}
\label{eq:staticquadrupoletides}
    Q_{ab}^{(i)}=-m_i^5\Lambda_2^{(i)} {\cal E}_{ab}\,.
\end{align}
We then introduce the effective tidal parameter~$\tilde{\Lambda}$ defined by~\cite{Favata:2013rwa}\footnote{Note that Ref.~\cite{Favata:2013rwa} adopts the convention $m_2\ge m_1$, opposite to ours in Eq.~\eqref{eq:massratioq}, resulting in a sign reversal of the terms proportional to the mass difference (see Eq.~(5) therein).}
\begin{equation}
\begin{aligned}
\label{LT}
  \tilde{\Lambda}=& \frac{8}{13}\left[\left(1+7\eta-31\eta^2\right)\left(\Lambda_2^{(1)}+\Lambda_2^{(2)}\right) \right.\\
  &\left.+\delta_{\eta}\left(1+9\eta-11\eta^2\right)\left(\Lambda_2^{(1)}-\Lambda_2^{(2)}\right)\right]. \end{aligned}
\end{equation}
The resulting expression for ${\cal A}_5$ reads
\begin{equation}
\label{eq:A5}
\mathcal{A}_{5}
=
-\frac{39}{2}\tilde{\Lambda}\,.
\end{equation}
For the equal-mass case, for which $\eta=1/4$, $\delta_\eta =0$, and $\Lambda_2^{(1)}=\Lambda_2^{(2)}$, Eq.~\eqref{LT} reduces to $\tilde{\Lambda}=\Lambda_2^{(1)}=\Lambda_2^{(2)}$.

\subsubsection{$6$PN term}
The $6$PN tidal contribution contains the next-to-leading-order correction to the static electric quadrupolar tidal contribution characterized by Eq.~\eqref{eq:staticquadrupoletides} and the leading-order magnetic quadrupolar tidal contribution. The magnetic response is characterized by the dimensionless magnetic deformability~$\Sigma_2^{(i)}$ defined through the linear relation~\cite{Damour:2009vw, Binnington:2009bb}
\begin{align}
S_{ab}^{(i)}=m_i^5 \Sigma_2^{(i)}{\cal B}_{ab},
\end{align}
where $S_{ab}^{(i)}$ and ${\cal B}_{ab}$ denote the induced current quadrupole moment and the quadrupolar magnetic tidal field, respectively.

The corresponding PN coefficient~${\cal A}_6$ is given by~\cite{Favata:2013rwa,JimenezForteza:2018rwr}
\begin{equation}
\label{eq:A6}
\mathcal{A}_{6}
=
-\frac{3115}{64}\tilde{\Lambda}
+
\frac{6595}{364}\delta_{\eta}\,\delta\tilde{\Lambda}+\tilde{\Sigma},
\end{equation}
where
\begin{equation}
\begin{aligned}
\label{dLT}
&\delta\tilde{\Lambda}=\frac{1}{2}\left[\delta_{\eta}\left(1-\frac{13272}{1319}\eta+\frac{8944}{1319}\eta^2\right)\left(\Lambda_2^{(1)}+\Lambda_2^{(2)}\right)\right.\\
&+\left.\left(1-\frac{15910}{1319}\eta+\frac{32850}{1319}\eta^2+\frac{3380}{1319}\eta^3\right)\left(\Lambda_2^{(1)}-\Lambda_2^{(2)}\right)\right].
\end{aligned}
\end{equation}
and
\begin{equation}
    \label{st}
    \tilde{\Sigma}=\left(\frac{6920}{7}-\frac{20740}{21X_1}\right)X_1^5\Sigma_2^{(1)}+\left(\frac{6920}{7}-\frac{20740}{21X_2}\right)X_2^5\Sigma_2^{(2)}.
\end{equation}
For an equal-mass case with $\Lambda_2^{(1)}=\Lambda_2^{(2)}$ and $\Sigma_2^{(1)}=\Sigma_2^{(2)}$, we have $\eta=1/4$ and $\delta_\eta=0$, and hence, $\delta \tilde{\Lambda}=0$, while Eq.~\eqref{st} reduces to $\tilde{\Sigma}=-(1295/21) \Sigma_2^{(1)}=-(1295/21) \Sigma_2^{(2)}$. Note that ${\rm sgn}(\tilde{\Sigma})=-{\rm sgn}(\Sigma)$ for $X_1, X_2 \le 1037/1038 (\simeq 0.999)$. The sign of $\Sigma_2^{(i)}$ depends on the assumed state of the perturbed fluid: $\Sigma_2^{(i)}$ is positive for strictly static fluids, whereas it is negative for irrotational fluids~\cite{Landry:2015cva}. The former corresponds to an isolated point in the solution space, whereas the latter arises as the zero-frequency limit of frequency-dependent perturbations and is therefore regarded as the physically relevant configuration for binary inspirals~\cite{Pani:2018inf}.

In the numerical implementation used to perform the analyses discussed in Sec.~\ref{sec:results}, the magnetic quadrupolar tidal deformabilities are not sampled independently. Indeed, ~$\Sigma_2^{(i)}$ are determined through the approximately EoS-insensitive quasi-universal relations~(QURs) of Refs.~\cite{Yagi:2013awa,JimenezForteza:2018rwr,Pani:2018inf}. Specifically, assuming the physically relevant stellar configurations in inspirals, we use a power-law fit for $\Sigma_2^{(i)}$ for the irrotational state,
\begin{equation}
\begin{aligned}
\label{eq:magneticQUR}
&\ln\left(-\Sigma_2^{(i)}\right)=-2.03+4.87\times 10^{-1}\left(\ln \Lambda_2^{(i)}\right)\\
&+9.69\times 10^{-3}\left(\ln \Lambda_2^{(i)}\right)^2+1.03\times 10^{-3}\left(\ln \Lambda_2^{(i)}\right)^3\\
&-9.37\times 10^{-5}\left(\ln \Lambda_2^{(i)}\right)^4+2.24\times 10^{-6}\left(\ln \Lambda_2^{(i)}\right)^5.
\end{aligned}
\end{equation}
This prescription is applied consistently to both the injected and recovery waveforms.

\subsubsection{$6.5$PN term}

The $6.5$PN tidal contribution contains the tidal-tail term and spin-tidal couplings. At this order, the general tidal phase also contains contributions involving rotational tidal Love numbers~\cite{Pani:2015nua,Abdelsalhin:2018reg}, which we denote collectively by $\hat{\gamma}_T$.

With this prescription, the $6.5$PN coefficient implemented in the waveform phase takes the form~\cite{Abdelsalhin:2018reg}\footnote{Note the differences in conventions with respect to Eq.~(15) of Abdelsalhin, Gualtieri, and Pani (AGP)~\cite{Abdelsalhin:2018reg}: $\tilde{\Lambda}=-(2/39)\Lambda_{\rm AGP}$, $\hat{\lambda}_T=\tilde{\Lambda}_{\rm AGP}$, $\hat{\sigma}_T=\hat{\Sigma}_{\rm AGP}$, and $\hat{\gamma}_T=\tilde{\Gamma}_{\rm AGP}$. Furthermore, our definition of $\Sigma_2^{(i)}$ differs from that of $\Sigma$ in Eq.~(18) of Ref.~\cite{Abdelsalhin:2018reg}.}
\begin{equation}
\mathcal{A}_{13/2}
=
\frac{39\pi}{2}\tilde{\Lambda}
+
\hat{\lambda}_T
+
\hat{\sigma}_T
+
\hat{\gamma}_T
.
\label{eq:A65PN}
\end{equation}
Here, the first term represents the tidal-tail coupling. The second term~$\hat{\lambda}_T$ denotes the coupling of the spins to the static electric quadrupolar tidal contribution characterized by Eq.~\eqref{eq:staticquadrupoletides}, while $\hat{\sigma}_T$ describes the corresponding coupling to the magnetic quadrupolar tidal contribution.

The parameters~$\hat{\lambda}_T$ and $\hat{\sigma}_T$ read
\begin{align}
\hat{\lambda}_T
={}&
X_1^5\Lambda_2^{(1)}
\bigg[
\left(
\frac{593}{4}
-\frac{1105}{8X_1}
+\frac{567X_1}{8}
-81X_1^2
\right)\chi_2
\label{eq:lambda_hat_65}\\
&+
\left(
-\frac{6607}{8}
+\frac{6639X_1}{8}
-81X_1^2
\right)\chi_1
\bigg]
+(1\leftrightarrow2),
\nonumber
\\[1ex]
\hat{\sigma}_T
={}&
X_1^5\Sigma_2^{(1)}
\bigg[
\left(
-\frac{9865}{3}
+\frac{4933}{3X_1}
+1644X_1
\right)\chi_2
-\chi_1
\bigg]\nonumber\\
&+(1\leftrightarrow2),
\label{eq:sigma_hat_65}\\[1ex]
\hat{\gamma}_T={}&
\frac{\chi_1}{M^4}\bigg[(856X_1-816X_1^2)\lambda_{23}^{(1)}-\left(\frac{833X_1}{3}-278X_1^2\right)\sigma_{23}^{(1)}\nonumber\\
&-
\eta\left(272\lambda_{32}^{(1)}-204\sigma_{32}^{(1)}\right)\bigg]+(1\leftrightarrow2),
\label{eq:gamma_hat_65}
\end{align}
where $\chi_i$ denotes the dimensionless spin of the $i$-th component. The notation~$(1\leftrightarrow2)$ represents the additional term obtained by exchanging all component labels.

For the nonspinning injections considered in this work, all spin-tidal couplings vanish, $\hat{\lambda}_T=\hat{\sigma}_T=\hat{\gamma}_T=0$. In the recoveries, $\hat{\lambda}_T$ and $\hat{\sigma}_T$ are retained and can be nonzero, since the component spins are sampled. In contrast, we set all rotational tidal Love numbers to zero, and hence $\hat{\gamma}_T=0$, also in the recoveries. 
This choice is justified because $\hat{\gamma}_T$ is linear in the component spins [see Eq.~\eqref{eq:gamma_hat_65}], which are restricted to $|\chi_i|\le0.2$ by our prior (see Appendix~\ref{app:priors}) and are constrained to $|\chi_i|\lesssim0.1$ already at $\text{SNR}=50$ (see Fig.~\ref{fig:corner_full_M1_snr50}), and are measured even more precisely at higher SNR. The inclusion of rotational tidal Love numbers is left to future work on spinning BNSs.

\subsubsection{$7$PN term}
At 7PN order, we include two tidal effects: the next-to-next-to-leading-order correction to the static electric quadrupolar tidal term and the leading-order static electric octupolar tidal term~\cite{Abdelsalhin:2018reg, Henry:2020ski,Chakraborty:2026qru}. The former is characterized by Eq.~\eqref{eq:staticquadrupoletides}, while the dimensionless electric octupolar tidal deformability~\(\Lambda_{3}^{(i)}\) is defined through the linear relation between the induced mass octupole moment~$Q_{abc}^{(i)}$ and the octupolar electric tidal field~${\cal E}_{abc}$~\cite{Binnington:2009bb}
\begin{align}
\label{eq:staticoctipoletides}
    Q_{abc}^{(i)}=-m_i^7 \Lambda_3^{(i)}{\cal E}_{abc}.
\end{align}
The corresponding PN coefficient~${\cal A}_7$ reads\footnote{Note the differences in conventions with respect to Eq.~(D1) of Ref.~\cite{Abdelsalhin:2018reg}: $\Lambda_3^{(i)}=(\lambda_3^{(i)})_{\rm AGP}/m_i^7$ and $X_i=(\eta_i)_{\rm AGP}$.}
\begin{equation}
\begin{split}
\mathcal{A}_{7}
=
\frac{128\eta}{3}
\Bigg\{
&
C_{\tilde{\Lambda}}\,\tilde{\Lambda}
+
C_{\delta\tilde{\Lambda}}\,
\delta\tilde{\Lambda}
-
\frac{125}{12}
\left[
X_1^5\Lambda_{3}^{(1)}
+
X_2^5\Lambda_{3}^{(2)}
\right]
\Bigg\},
\end{split}
\label{eq:A7PN}
\end{equation}
where
\begin{equation}
\begin{aligned}
    C_{\tilde{\Lambda}}=& -\frac{5}{1820786688 \cdot\eta\left(53-2738\eta\right)}\\
    &
    \times \left(103964343593-4517921513750\eta\right.\\
    &\quad \left.+2508748099960\eta^2-1486330141504\eta^3\right),\\
    C_{\delta\tilde{\Lambda}}=& \frac{6595 \delta_\eta}{246564864\cdot \eta \left(53-2738\eta\right)} \\
    &\times \left(8394865-261503638\eta-25585568\eta^2\right).
\end{aligned}
\end{equation}
The octupolar tidal deformabilities are determined from the corresponding quadrupolar tidal deformabilities using the quasi-universal multipole Love relation of~\cite{Yagi:2013sva},
\begin{equation}
\begin{aligned}
\ln \Lambda_3^{(i)}
={}& -1.15
+1.18\ln\Lambda_2^{(i)}
+2.51\times10^{-2}\left(\ln\Lambda_2^{(i)}\right)^2\\
&-1.31\times10^{-3}\left(\ln\Lambda_2^{(i)}\right)^3
+2.52\times10^{-5}\left(\ln\Lambda_2^{(i)}\right)^4.
\end{aligned}
\label{eq:lambda3_qur}
\end{equation}

\subsubsection{$8$PN term}
At 8PN order, we include two physically distinct contributions:  the leading dynamical tidal contribution~\cite{Apostolidis:2026qsg} and the static quadratic tidal contribution~\cite{Pani:2025qxs}. The former corresponds to the next-to-leading-order conservative correction to the linear tidal response and arises at second order in a low-frequency (or time derivative) expansion of the tidal perturbation~\cite{Poisson:2020vap,Pitre:2023xsr,HegadeKR:2024agt}. The latter arises from the quadratic coupling to external tidal fields and is analyzed within second-order tidal perturbation theory~\cite{Poisson:2020vap,Pitre:2025qdf}. 

For the $i$-th star, the dynamical and quadratic contributions to the induced mass quadrupole moment read\footnote{The parameter~$\Lambda_{\rm dyn}^{(i)}$ is identical to $\Lambda_2$ defined in Eq.~(59) of Ref.~\cite{Apostolidis:2026qsg}. Note that $\bar{\zeta}^{(i)}$ is defined as a dimensionless parameter with the opposite sign to the dimensionless normalization of $\lambda_{222}^{(i)}$ introduced in Eq.~(3) of Ref.~\cite{Pani:2025qxs}:
\begin{equation}
\bar{\zeta}^{(i)}
=
-\frac{G\lambda_{222}^{(i)}}{(Gm_i)^7}>0.
\label{eq:zeta_def}
\end{equation}}
\begin{equation}
Q_{ab}^{(i)}
=
\frac{\left( G m_i\right)^7}{G}\Lambda_{\rm dyn}^{(i)}\ddot{\cal E}_{ab}+6\frac{\left(G m_i\right)^7}{G}\bar{\zeta}^{(i)}
\mathcal{E}_{\langle a}^{\ \ c}
\mathcal{E}_{b\rangle c},
\label{eq:quadratic_response}
\end{equation}
where the double dots denote second time derivatives. Here, the first term is associated with the dynamical response characterized by the dimensionless dynamical tidal deformability~$\Lambda_{\rm dyn}^{(i)}$, while the second term describes the quadratic response characterized by the dimensionless quadratic tidal deformability~$\bar{\zeta}^{(i)}$. The dynamical tidal deformability~$\Lambda_{\rm dyn}^{(i)}$ exhibits logarithmic running, and therefore depends on the binary scale, schematically, in the form of~\cite{Apostolidis:2026qsg} 
\begin{align}
    \Lambda_{\rm dyn}^{(i)}= \hat{\Lambda}_{\rm dyn}^{(i)}-\beta_2 ^{(i)}\ln\left(2Gm_i \mu \right),
\end{align}
where $\mu$ is the renormalization scale, and $ \hat{\Lambda}_{\rm dyn}^{(i)}$ denotes the scale-independent part of $\Lambda_{\rm dyn}^{(i)}$. The parameter~$\beta_2$ encodes the logarithmic running and is defined by
\begin{equation}
\beta_2^{(i)}
=
-\frac{64}{45}
\left(
1+\frac{321}{112}\Lambda_2^{(i)}
\right).
\label{eq:beta2_dyn}
\end{equation}

The $8$PN coefficient~${\cal A}_8$ is given by~\cite{Pani:2025qxs,Apostolidis:2026qsg}
\begin{equation}
\begin{aligned}
\label{eq:A8_total}
\mathcal{A}_{8}
=&- \frac{15}{11} \sum_{i=1}^2 X_i^5\Bigg[
 \left(8X_i+147 \eta \right)\hat{\Lambda}_{\rm dyn}^{(i)} \\
&+\left(\frac{38}{11}X_i +\frac{1253}{44}\eta\right)\beta_2^{(i)}\\
&+
\left(
8X_i+147\eta
\right)
\beta_2^{(i)}
\ln\left(
\frac{1}{2 x X_i}
\right)\Bigg]\\
&-\frac{45}{11}
\left(
19\tilde{\zeta}_{+}
-
17\delta_{\eta}\,\tilde{\zeta}_{-}
\right),
\end{aligned}
\end{equation}
where the dimensionless combinations $\tilde{\zeta}_{\pm}$ are defined by
\begin{equation}
\tilde{\zeta}_{\pm}
=
2
\left[
X_1^6X_2\,\bar{\zeta}^{(1)}\pm
X_2^6X_1\,\bar{\zeta}^{(2)}
\right],
\label{eq:zeta_pm}
\end{equation} 
Following Ref.~\cite{Apostolidis:2026qsg}, we identify the renormalization scale with the inverse orbital separation, $\mu=1/r$. At leading PN order, $r=M/x$ in geometrized units, so that $2m_i\mu=2X_i x$. 

In the waveform implementation, \(\bar{\zeta}^{(i)}\) and $\hat{\Lambda}_{\rm dyn}^{(i)}$ are not treated as independent parameters; instead, they are expressed in terms of the static electric quadrupolar deformability~\(\Lambda_2^{(i)}\) using the QURs~\cite{Pani:2025qxs,Apostolidis:2026qsg}\footnote{{The coefficient~$\lambda_2$ in Eq.~(77) of Ref.~\cite{Pani:2025qxs} differs in normalization from the dimensional deformability~$m_i^5 \Lambda_2^{(i)}$ defined in Eq.~\eqref{eq:staticquadrupoletides} of the present work with $m_i^5 \Lambda_2^{(i)}= 4 \lambda_2^{(i)}|_{\text{Ref.~\cite{Pani:2025qxs}}}$~(see also footnote~13 in the reference).  Also note that Eq.~(4.4) of Ref.~\cite{Mandal:2024iug} appears to require an overall factor of $1/2$, so that $\tilde{\lambda}_{(+)}=\tilde{\lambda}_{(1)}=\tilde{\lambda}_{(2)}$ and $\tilde{\lambda}_{(-)}=0$ when the two objects in the binary are identical, as stated in the reference.}}
\begin{equation}
\ln\bar{\zeta}^{(i)}
=
3.77
+
1.51\left[\ln\left(\dfrac{\Lambda_2^{(i)}}{4}\right)\right]
+
0.005\left[\ln\left(\dfrac{\Lambda_2^{(i)}}{4}\right)\right]^2,
\label{eq:qur_quad}
\end{equation}
and 
\begin{equation}
\ln\hat{\Lambda}_{\rm dyn}^{(i)}
=
3.3840
+
1.1621\left(\ln\Lambda_2^{(i)}\right)
+
0.019180\left(\ln\Lambda_2^{(i)}\right)^2.
\label{eq:qur_dyn}
\end{equation}

Although the EoS-insensitivity of these relations is known to deteriorate towards small compactness~\cite{Sham:2014kea,Silva:2016myw}, the QURs in Eqs.~\eqref{eq:lambda3_qur}, \eqref{eq:qur_quad}, and~\eqref{eq:qur_dyn} remain accurate up to $\Lambda_2\sim10^5$~\cite{Yagi:2013sva,Pani:2025qxs,Apostolidis:2026qsg,Chakraborty:2026qru}, which covers the range of tidal deformabilities relevant for the subsolar-mass binaries considered here. The relations for $\kappa_i$ and $\Sigma_2^{(i)}$, Eqs.~\eqref{eq:loveQ} and~\eqref{eq:magneticQUR}, have a negligible impact on our results, since the injections are nonspinning and the magnetic contribution is subdominant.

\subsection{Reference and cutoff frequencies}
\label{subsec:frequencies}

The higher-order tidal corrections considered in this work become increasingly important during the late inspiral. Consequently, assessing the impact of the 7PN and 8PN terms requires a careful choice of the frequency range over which the PN waveform is assumed to remain valid. Since \texttt{TaylorF2} describes only the inspiral, extrapolating it arbitrarily close to merger could artificially enhance higher-order tidal contributions in a regime where the PN expansion becomes increasingly unreliable.

We address this issue by terminating the waveform at a physically motivated cutoff frequency. For each binary, we define the source-frame cutoff frequency as
\begin{equation}
f_{\rm cut}^{\rm src}
=
\min\left(
f_{\rm Roche}^{\rm src},
f_{\rm ISCO}^{\rm src}
\right),
\label{eq:fcut}
\end{equation}
where $f^{\rm src}_{\rm Roche}$ estimates the onset of tidal disruption, while $f^{\rm src}_{\rm ISCO}$ denotes the GW frequency at the Schwarzschild innermost stable circular orbit,
\begin{equation}
f^{\rm src}_{\rm ISCO}
=
\frac{c^3}
{6^{3/2}\pi G M}
\simeq
4.4\,{\rm kHz}
\left(
\frac{M_\odot}{M}
\right).
\label{eq:fisco}
\end{equation}
The Roche frequency is computed in the source frame using the empirical fit of Ref.~\cite{Bandopadhyay:2022tbi}\footnote{We use the data available at \href{https://github.com/sugwg/sub-solar-ns-detectability}{https://github.com/sugwg/sub-solar-ns-detectability}.}
\begin{equation}
\begin{split}
\frac{f_{\rm Roche}^{\rm src}}{\mathrm{Hz}}
=&\,
-26.9
-35.5
\left(
\frac{m_1}{M_\odot}
\right)
-3.02
\left(
\frac{m_1}{M_\odot}
\right)^2
\\
&+
1690
\left(
\frac{m_2}{M_\odot}
\right)
-
575
\left(
\frac{m_2}{M_\odot}
\right)^2\,,
\end{split}
\label{eq:froche}
\end{equation}
which was calibrated against numerical quasi-equilibrium sequences for component masses $m_i\in[0.2,1]\,M_\odot$, reproducing the results for the APR EoS with errors below $10\%$~\cite{Bandopadhyay:2022tbi,Crescimbeni:2024cwh}. Its application to binaries with components outside this mass range therefore involves an extrapolation, the accuracy of which is not guaranteed. As a consistency check, a Roche-lobe estimate based on Eggleton's approximation ~\cite{eggleton1983approximations} and the APR mass--radius relation yields $f_{\rm Roche}\simeq 1.15\,\mathrm{kHz}$ for $M_{\rm tot}=2.8\,M_\odot$ and $q=0.93$, comparable to the extrapolated fit. We thus adopt this prescription as an approximate waveform termination criterion, while acknowledging the uncertainty associated with its extrapolation.

The corresponding detector-frame frequencies are related to their source-frame values by
\begin{equation}
f
=
\frac{f^{\rm src}}{1+z},
\end{equation}
where $z$ is the cosmological redshift.

Rather than imposing a sharp cutoff, which would introduce an artificial discontinuity in the frequency-domain waveform, we smoothly suppress the signal around $f_{\rm cut}$ using the phenomenological tapering function~\cite{Maselli:2013rza,DeLuca:2022xlz}
\begin{equation}
T(f)
=
\frac{
1+\exp\!\left(-\lambda_f/\Delta\lambda_f\right)
}{
1+\exp\!\left[
\left(f/f_{\rm ISCO}-\lambda_f\right)
/\Delta\lambda_f
\right]
},
\label{eq:taper}
\end{equation}
which depends on the dimensionless cutoff parameter
\begin{equation}
\lambda_f
=
\frac{f_{\rm cut}}{f_{\rm ISCO}},
\end{equation}
and on $\Delta\lambda_f$, which regulates the width of the transition. The waveform entering the likelihood is then
\begin{equation}
\tilde h_{\rm tapered}(f)
=
T(f)\,\tilde h(f).
\end{equation}
All frequencies entering the taper are evaluated in the detector frame. The numerical frequency grid extends beyond the nominal cutoff only sufficiently far to capture the exponentially suppressed tail of the waveform and never exceeds $f_{\rm ISCO}$.

\subsection{Relative importance of tidal corrections}
\label{subsec:trends}

\begin{figure*}[t]
    \centering
    \includegraphics[width=0.49\textwidth]{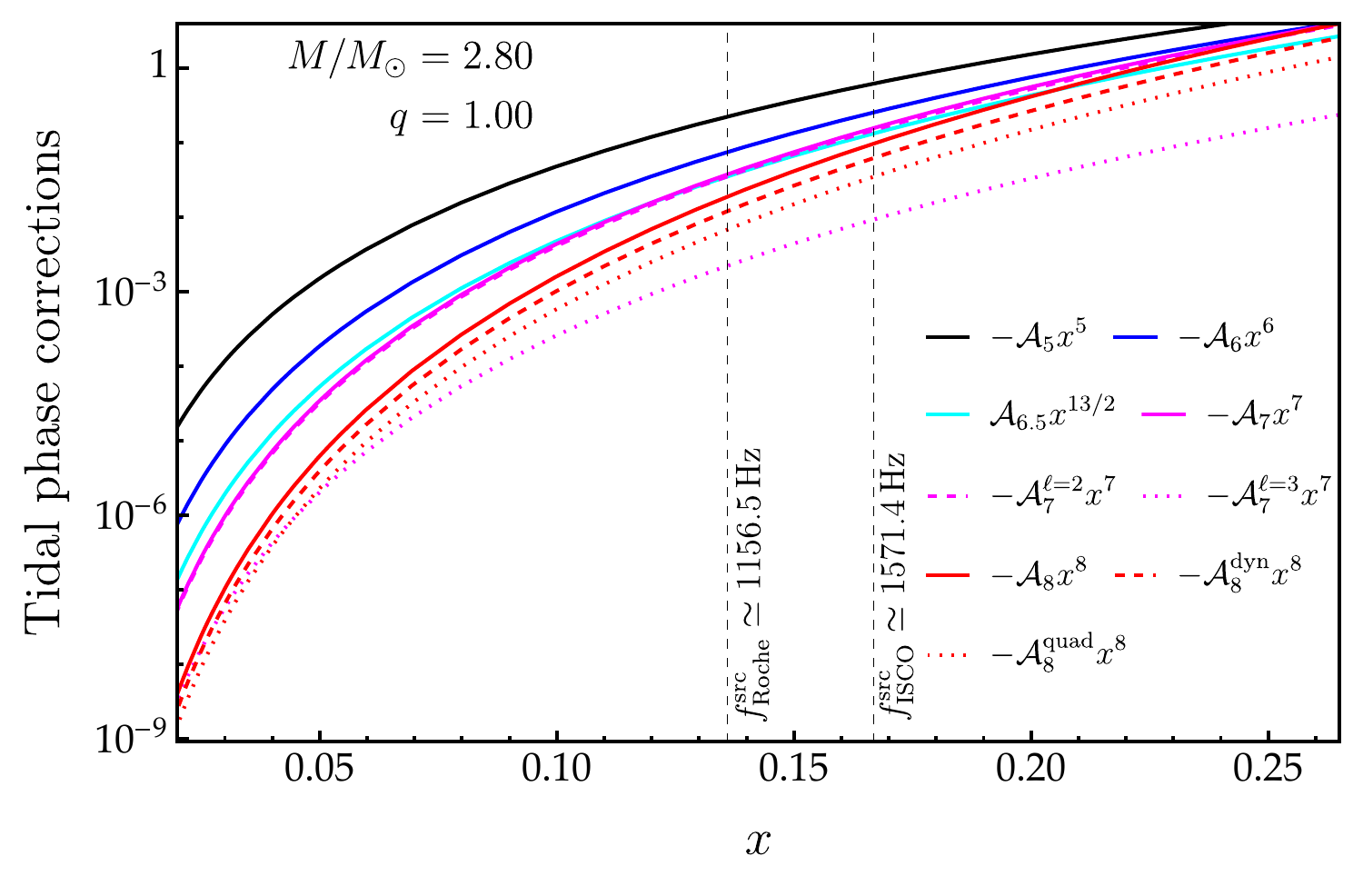}
    \includegraphics[width=0.49\textwidth]{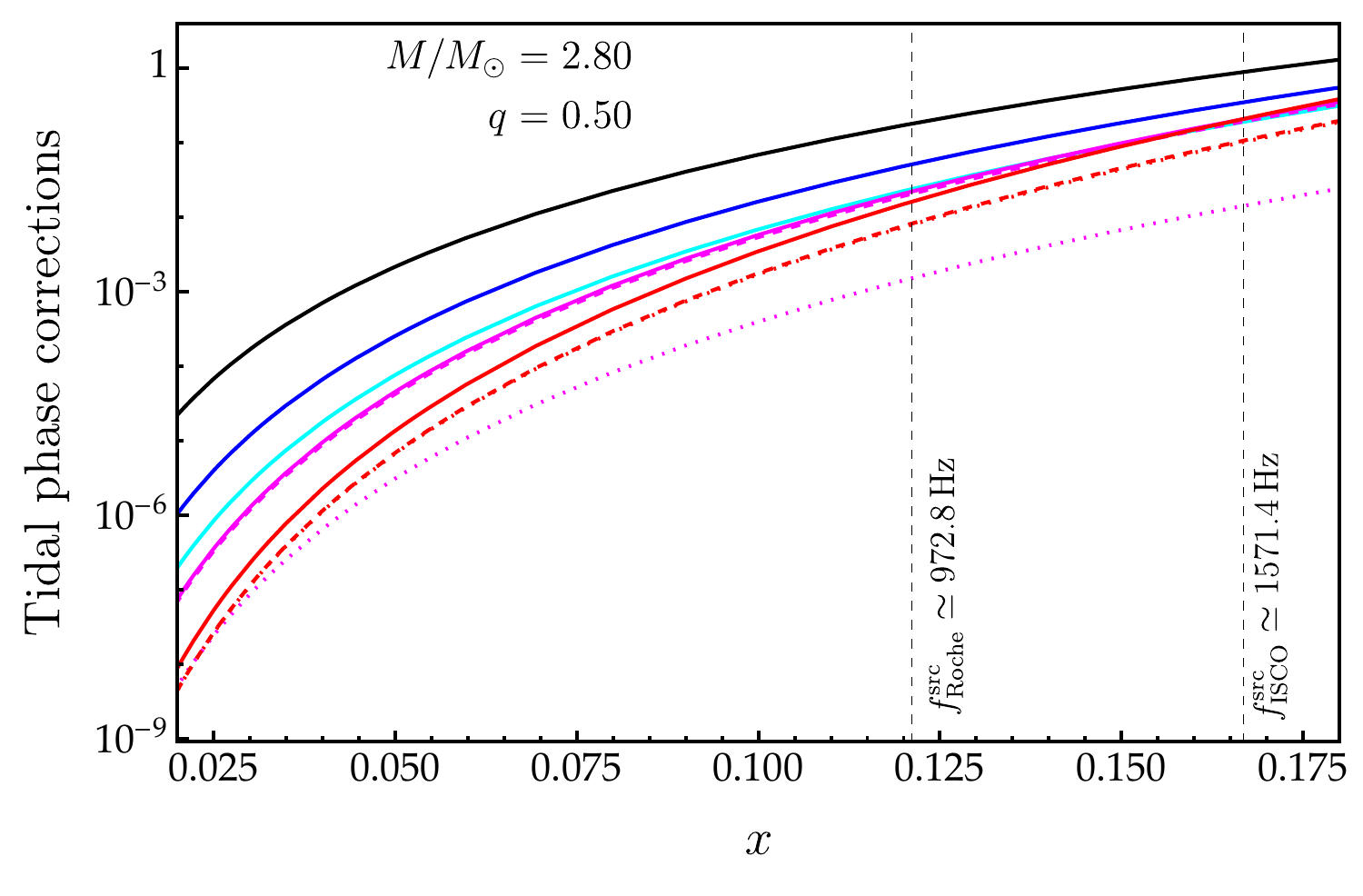}
      \includegraphics[width=0.49\textwidth]{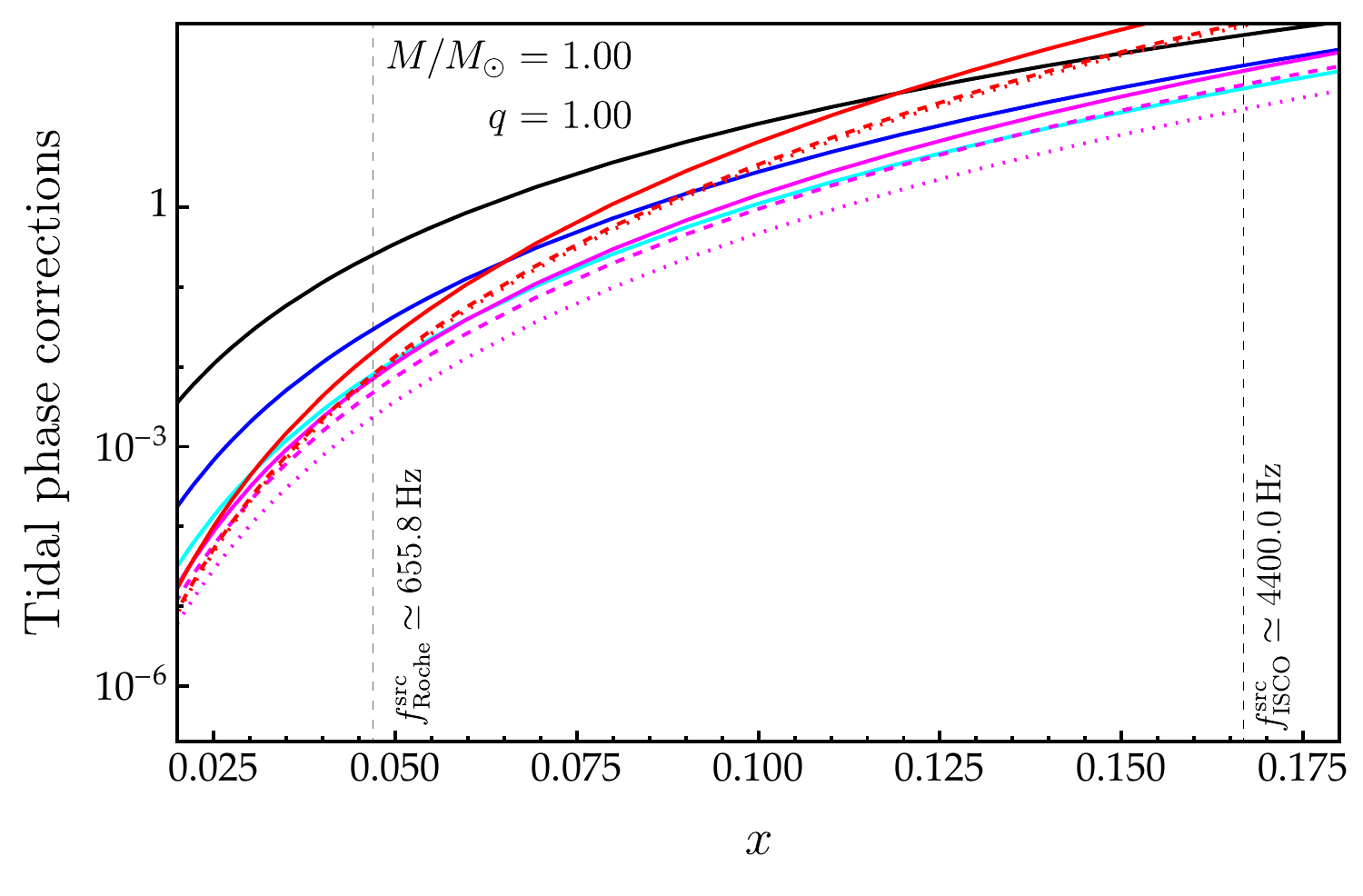}
    \includegraphics[width=0.49\textwidth]{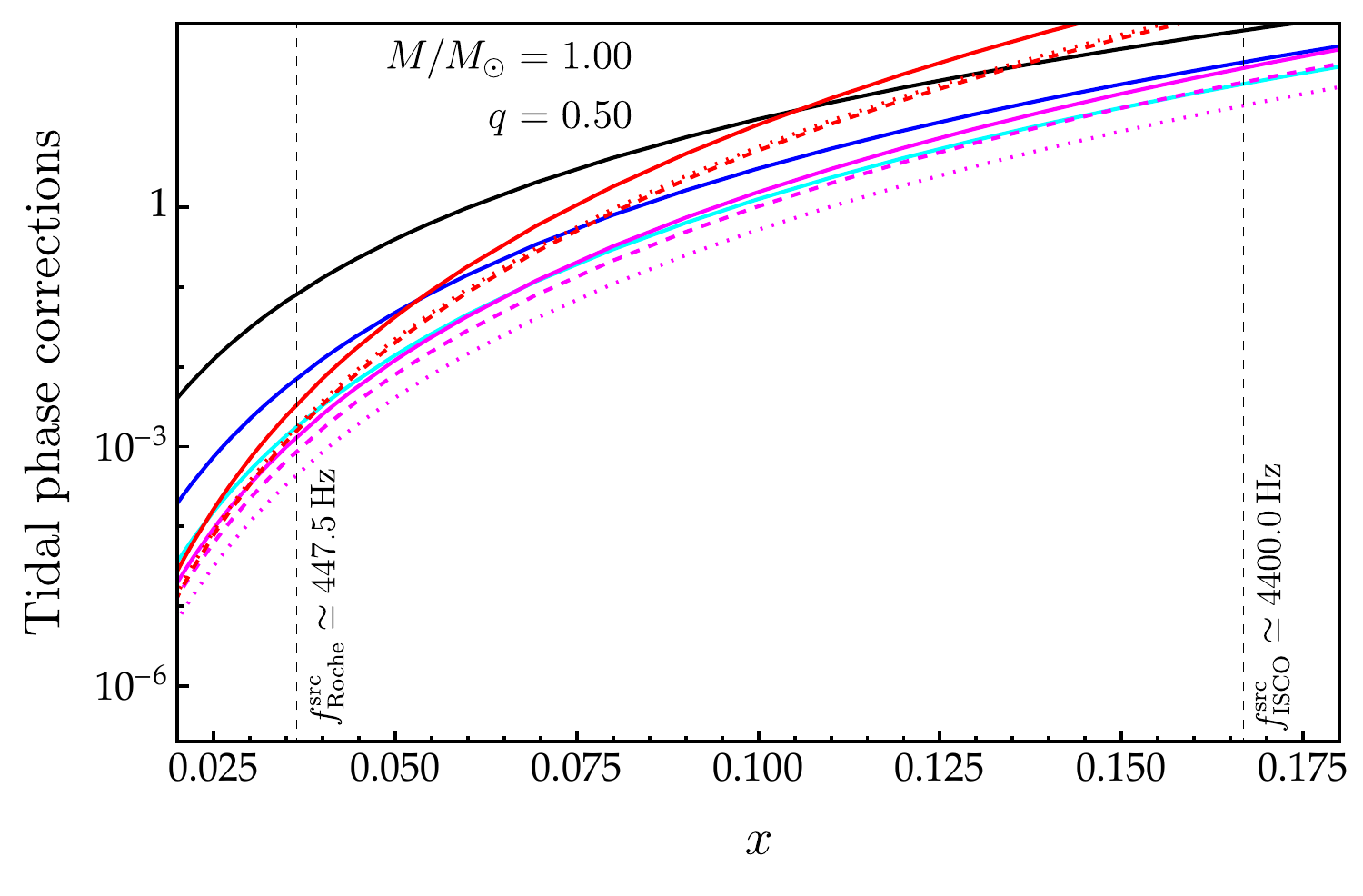}
    \caption{Comparison of the absolute value of the individual tidal phase contributions to $(128/3)\eta x^{5/2}\delta\psi_{\rm tidal}$, given in Eq.~\eqref{eq:tidal_full_8PN}, at different PN orders as functions of $x=(M \omega)^{2/3}=(\pi f^{\rm src} M)^{2/3}$. The APR EoS is assumed for reference.
    }
    \label{fig:phase}
\end{figure*}

Prior to performing parameter estimation with the waveform approximant introduced above, we study the individual tidal contributions to $\delta \psi_{\rm tidal}$, given in Eq.~\eqref{eq:tidal_full_8PN}, at different PN orders. Figure \ref{fig:phase} shows the individual terms in Eq.~\eqref{eq:tidal_full_8PN}, corresponding to the $i$-th PN-order tidal corrections, as functions of $x$, for different choices of the total mass and mass ratio. As a general trend, $f_{\rm Roche}$ decreases as total mass and/or mass ratio decrease, while $f_{\rm ISCO}$ increases with decreasing total mass and is independent of mass ratio, as evident from Eq.~\eqref{eq:fisco}. The magnitudes of all contributions increase with $x$ and have the same sign, except for the tidal-tail contribution at $6.5$PN. 

At $6$PN order, setting $\tilde{\Sigma}=0$ produces no appreciable change in the results, indicating that the next-to-leading-order correction to the static electric quadrupolar tidal contribution dominates over the static magnetic quadrupolar contribution. This hierarchy arises because $\Sigma_2^{(i)}$ is typically about two orders of magnitude smaller than $\Lambda_2^{(i)}$, as can be directly inferred from the QUR in Eq.~\eqref{eq:magneticQUR}. Our findings are consistent with those of Refs.~\cite{JimenezForteza:2018rwr,Gupta:2023oyy}.

At 7PN order, the quadrupole contribution is generically larger than the octupole contribution, while the difference becomes smaller for lower total mass. This is because the dimensionless tidal deformabilities scale with the stellar compactness~$C_{(i)}$ as $\Lambda_\ell^{(i)}\propto k_\ell^{(i)}C_{(i)}^{-(2\ell+1)}$ for the $\ell$th electric-type tidal Love numbers~$k_\ell^{(i)}$. In particular, $\Lambda_3^{(i)}/\Lambda_2^{(i)}=(1/5)(k_3^{(i)}/k_2^{(i)})C_{(i)}^{-2}$. Since the ratio~$k_3^{(i)}/k_2^{(i)}$ are of comparable magnitude and typically varies more mildly than the explicit compactness factor~\cite{Binnington:2009bb,Damour:2009vw,Katagiri:2025qze}, the octupole contribution becomes relatively more important as the stellar compactness decreases, modulo the EoS dependence of the tidal Love numbers. 

At $8$PN order, the dynamical tidal contribution is generically larger than the quadratic tidal contribution in the cases considered, although the latter can become larger for sufficiently low total mass and mass ratio. This trend can be understood in terms of the induced mass quadrupole moment given in Eq.~\eqref{eq:quadratic_response}. For the second star in a quasi-circular binary, the tidal-field contributions sourced by the first star and entering the quadratic and dynamical responses are related as~\cite{Pitre:2023xsr} 
\begin{equation}
\quad \mathcal{E}_{\langle a}^{\ \ c}
\mathcal{E}_{b\rangle c} \sim  \frac{1}{1+q} \ddot{\cal E}_{ab}.
\end{equation}
The relative magnitude of the quadratic and dynamical responses of the second star therefore scales schematically as
\begin{align}
\label{eq:ratiodynvsquadr}
    \frac{\rm \#quadratic}{\rm \#dynamical}\sim \frac{\bar{\zeta}^{(2)}}{\Lambda_{\rm dyn}^{(2)}}\frac{1}{1+q}.
\end{align}
The total-mass dependence is encoded in $\bar{\zeta}^{(2)}/\Lambda_{\rm dyn}^{(2)}$. In particular, combining the results of Refs.~\cite{Pani:2025qxs,Apostolidis:2026qsg}, the ratio~$\bar{\zeta}^{(i)}/\Lambda_{\rm dyn}^{(i)}$ increases towards lower compactness and can exceed unity, as also implied by the QURs~\eqref{eq:qur_quad} and~\eqref{eq:qur_dyn}. Moreover, decreasing $q$ increases $1/(1+q)$, directly enhancing the quadratic response of the lighter star relative to the dynamical response.

Lower-PN-order contributions generically dominate over higher-order contributions at sufficiently small~$x$. However, the difference between the $6.5$PN and $7$PN contributions is relatively small, and as $x$ increases, the latter eventually exceeds the former. This suggests that the $7$PN contribution, particularly the next-to-next-to-leading-order electric quadrupolar tidal contribution, could become comparable to, or even exceed, the $6.5$PN contribution during the late inspiral. The $8$PN contribution can also become larger than the $7$PN contribution. The corresponding crossover frequencies decrease as the total mass and/or mass ratio decrease. In particular, for the typical equal-mass binary~$(M/M_\odot,q)=(2.8,1)$, the crossing between the $6.5$PN and $7$PN contributions occurs below $f_{\rm Roche}$, implying the relative importance of the $7$PN contribution even for standard binaries. For lower-total mass systems, the crossing between the $7$PN and $8$PN contributions also occurs below $f_{\rm Roche}$, suggesting that the $8$PN contribution could be relevant for parameter estimation of subsolar-mass BNSs.

The crossover frequencies can be analytically estimated as follows. For the crossing between the $6.5$PN and $7$PN contributions, equating the magnitudes of ${\cal A}_{6.5}x^{6.5}$ and ${\cal A}_7x^7$ evaluated at $q=1$, we obtain the approximate expression for the corresponding crossover frequency, 
\begin{align}
\label{eq:fc657}
    \frac{f_c^{(6.5,7)}}{\rm Hz}\simeq {\cal N}^{(6.5,7)}\left(\frac{M_\odot}{M}\right),
\end{align}
where
\begin{widetext}
    \begin{align}
    {\cal N}^{(6.5,7)}=\frac{3205}{\left(1+0.0132\tilde{\Lambda}^{0.18} e^{\left(\ln \tilde{\Lambda}\right)^2\left[0.0251-\left(0.00131-0.0000252\ln \tilde{\Lambda}\right)\ln \tilde{\Lambda}\right]}\right)^3}.
\end{align}
\end{widetext}
Here, ${\cal N}^{(6.5,7)}$ is a monotonically decreasing function of $\tilde{\Lambda}$, decreasing from ${\cal N}^{(6.5,7)}\simeq3000$ to ${\cal N}^{(6.5,7)}\simeq760$ over $10\le\tilde{\Lambda}\le 10^5$.
Likewise, by equating ${\cal A}_7x^7$ and ${\cal A}_8x^8$, with the coefficients evaluated at $q=1$ and $\beta_2^{(i)}=0$ so that the logarithmic running term in the dynamical tidal contribution is neglected, we obtain the approximate expression for the corresponding crossover frequency,
\begin{align}
\label{eq:fc78}
    \frac{f_c^{(7,8)}}{\rm Hz} \simeq {\cal N}^{(7,8)} \left(\frac{M_\odot}{M}\right),
\end{align}
with
\begin{widetext}
\begin{align}
    {\cal N}^{(7,8)}=\frac{134253}{\tilde{\Lambda}^{0.243} e^{0.0288 \left(\ln\tilde\Lambda\right)^2}}\left(\frac{1+0.0132\tilde{\Lambda}^{0.18} e^{\left(\ln \tilde{\Lambda}\right)^2\left[0.0251-\left(0.00131-0.0000252\ln \tilde{\Lambda}\right)\ln \tilde{\Lambda}\right]}}{1+0.128 \tilde{\Lambda}^{0.334} e^{-0.0142 \left(\ln\tilde \Lambda\right)^2}}\right)^{3/2},
\end{align}
\end{widetext}
which monotonically decreases with $\tilde{\Lambda}$ from ${\cal N}^{(7,8)}\simeq4.8\times10^4$ to ${\cal N}^{(7,8)}\simeq140$ over $10\le\tilde{\Lambda}\le 10^5$, i.e., much more steeply than ${\cal N}^{(6.5,7)}$.  Moreover, ${\cal N}^{(6.5,7)}>{\cal N}^{(7,8)}$ for $\tilde{\Lambda}\gtrsim2\times10^3$, implying that $f_c^{(7,8)} < f_c^{(6.5,7)}$ in the subsolar-mass regime.
This suggests an enhanced relative importance of the $8$PN-order correction in the subsolar-mass regime at low GW frequencies.


\section{Results}
\label{sec:results}
In this section, we present the main findings of this work. We first describe the Bayesian framework adopted here. We then assess the impact of the $7$PN and $8$PN tidal terms in the PN waveform using the EoS-insensitive QURs~\eqref{eq:qur_quad} and~\eqref{eq:qur_dyn}. Finally, we investigate the measurability of the $8$PN tidal contribution in an agnostic manner and study the estimation of the individual binary parameters,~$(m_1,m_2,\Lambda_2^{(1)},\Lambda_2^{(2)})$, both with and without imposing the QURs.

\subsection{Bayesian framework}

We consider mergers on quasicircular orbits which, in the BH case, are overall characterized by 11 parameters (see, e.g., ~\cite{Maggiore:2007ulw})
\begin{equation}\label{eq:paramBBH}
{\boldsymbol \theta}
= \{m_1, m_2, d_L, \iota, \alpha, \delta, \psi, t_c, \Phi_c, \chi_{1}, \chi_{2} \}\,, 
\end{equation}
where $\chi_{1,2}$ are the signed aligned-spin components; $\alpha$ and $\delta$ are the right ascension and declination, respectively; $\iota$ is the inclination angle of the binary with respect to the line of sight; $\psi$ is the polarization angle; $t_c$ the time at coalescence; and $\Phi_c$ the phase at coalescence.

To investigate the impact of higher-order terms in BNS waveforms, we perform parameter estimation and model comparison within the Bayesian framework. Given a set of source parameters $\boldsymbol{\theta}$, the detected GW data $d(t;\boldsymbol{\theta})$ is identified as:
\begin{equation}
d(t)
=
h(t;\boldsymbol{\theta})
+
n(t),
\end{equation}
where $h(t;\boldsymbol{\theta})$ is the GW signal and $n(t)$ denotes the detector noise. Bayes' theorem gives the posterior probability distribution
\begin{equation}
p(\boldsymbol{\theta}|d,\mathcal{H})
=
\frac{
\mathcal{L}(d|\boldsymbol{\theta},\mathcal{H})
\pi(\boldsymbol{\theta})
}{
\mathcal{Z}(d|\mathcal{H})
},
\label{eq:bayes}
\end{equation}
where $\mathcal{H}$ denotes the waveform model under consideration, $\mathcal{L}$ is the likelihood, $\pi$ is the prior distribution, and $\mathcal{Z}$ is the Bayesian evidence.

The prior $\pi(\boldsymbol{\theta}|\mathcal{H})$ and the posterior $p(\boldsymbol{\theta}|d,\mathcal{H})$ encode our knowledge of the source parameters $\boldsymbol{\theta}$ before and after observing the data $d$, respectively. The likelihood is determined by the waveform and the assumed noise model. Each detector is taken to have stationary Gaussian noise and the detectors collectively are assumed to be statistically independent.  The likelihood can therefore be factorized into the contributions from the individual instruments. Thus, the likelihood can be written as
\begin{equation}
\mathcal{L}(d|\boldsymbol{\theta}, \mathcal{H})
\propto \exp\left[
-\frac{1}{2}
\sum_{k}
\bigl(h_k(\boldsymbol{\theta})-d_k
\mid
h_k(\boldsymbol{\theta})-d_k\bigr)_k
\right],
\end{equation}
where $d_k$ and $h_k(\boldsymbol{\theta})$ denote, respectively, the data and the model waveform projected onto the $k$-th detector, and the sum is carried out over all detectors. The noise-weighted inner product for the $k$-th detector is defined as
\begin{equation}
(g\mid h)_k
=
4\,\mathrm{Re}
\int_{f_{\min}}^{f_{\max}}
\mathrm{d}f\,
\frac{\tilde{g}_k^{*}(f)\tilde{h}_k(f)}
{S_{n,k}(f)}\,,
\label{inner_product}
\end{equation}
where tildes denote Fourier transforms, and $S_{n,k}(f)$ is the one-sided noise power spectral density of the $k$-th detector.
Throughout this work, we set $f_{\min}=5\,\mathrm{Hz}$, while $f_{\max}$ is discussed in Subsec.~\ref{subsec:frequencies}.

The corresponding optimal matched-filter SNR is
\begin{equation}
\text{SNR}
=
\sqrt{
\sum_{k}
(h|h)_k
}\,.
\label{eq:network_snr}
\end{equation}
For concreteness, we consider a single ET observatory in its triangular configuration, comprising three nested interferometers with 10~km arms, each characterized by the corresponding noise power spectral density of Ref.~\cite{Branchesi:2023mws}. 
Nevertheless, we expect our results to depend primarily on the total SNR, rather than on the specific detector configuration.

In this work we analyze signals injected into zero-noise data. Therefore, the data entering the likelihood are given by the injected signal itself, $d=h_{\rm inj}$, while the detector sensitivity is still fully accounted for through the noise power spectral density in Eq.~\eqref{inner_product}. This choice removes fluctuations associated with a particular noise realization and allows us to isolate systematic differences between the waveform models.

Finally, we identify:
\begin{equation}
\mathcal{Z}(d|\mathcal{H})
=
\int
d\boldsymbol{\theta}\,
\mathcal{L}(d|\boldsymbol{\theta},\mathcal{H})
\pi(\boldsymbol{\theta}|\mathcal{H}),
\label{eq:evidence}
\end{equation}
as the evidence, which provides the quantity used for model comparison. Given two hypotheses (in our case, two waveform models) $\mathcal{H}_i$ and $\mathcal{H}_j$, we define the Bayes factor
\begin{equation}
\mathcal{B}_{ij}
=
\frac{
\mathcal{Z}(d|\mathcal{H}_i)
}{
\mathcal{Z}(d|\mathcal{H}_j)
},
\end{equation}
which is commonly expressed in $\log_{10}$ basis.
For reference, we adopt the commonly used Jeffreys scale~\cite{Jeffreys:1939xee}: $|\log_{10}\mathcal{B}_{ij}|<0.5$ is regarded as inconclusive, $0.5\leq|\log_{10}\mathcal{B}_{ij}|<1$ as substantial evidence, $1\leq|\log_{10}\mathcal{B}_{ij}|<2$ as strong evidence, and $|\log_{10}\mathcal{B}_{ij}|\geq2$ as decisive evidence.

We perform the Bayesian analysis using \textsc{Bilby}~\cite{Ashton:2018jfp} and evaluate the posterior distributions and evidences with the nested-sampling algorithm \textsc{Dynesty}~\cite{Speagle:2019ivv}. We reduce the computational cost of likelihood evaluations using the relative-binning technique, in which the waveform is expanded around a suitably chosen fiducial signal~\cite{Zackay:2018qdy, Narola:2023men}.

Typically, the parameters sampled in our analyses, that characterize BNSs subjected to tidal deformabilities, are~\cite{Flanagan:2007ix, Hinderer:2007mb, Wade:2014vqa, Dietrich_2019} 
\begin{equation}
\boldsymbol{\theta}_{\rm sampling}
=
\{
M_c,
\eta,
\chi_1,
\chi_2,
d_L,
\Phi_c,
t_c,
\tilde{\Lambda},
\delta\tilde{\Lambda},
\lambda_f
\},
\label{eq:sampled_parameters}
\end{equation}
where $\lambda_f=f_{\rm cut}/f_{\rm ISCO}$ parametrizes the waveform termination discussed in Sec.~\ref{subsec:frequencies}.
The sky position, polarization angle, and inclination are fixed to their injected values. This choice is motivated by two considerations. First, we adopt an optimistic observational scenario in which the sky position is determined through the detection of an electromagnetic counterpart, while the inclination can be constrained by the opening angle of an associated  jet~\cite{Gianfagna:2022kpw}. Second, fixing these parameters reduces correlations among the different parameters, allowing us to isolate the measurability and impact of the 7PN and 8PN tidal contributions, which are the primary focus of this work.

Allowing these parameters to vary is expected to introduce two main effects. First, the distance-inclination degeneracy would become relevant, although it can be partially broken by waveform precession and higher-order modes, or by combining observations from multiple detectors with different orientations, thereby improving the detector's polarization sensitivity~\cite{Usman:2018imj,Mascioli:2025cnw}. Second, the sky-localization posterior may exhibit multiple modes, particularly for a triangular ET configuration, as discussed in Refs.~\cite{Santoliquido:2026bej,Crescimbeni:2026dvg}. Nevertheless, we verified on a representative subset of injections that allowing these parameters to vary leaves the posterior distributions of the tidal parameters essentially unchanged, while the extrinsic parameters (sky position, distance, and inclination) remain poorly constrained by a single triangular ET. This suggests that our conclusions regarding the measurability of higher-order tidal effects are robust against dropping the simplifying assumptions adopted here.

\begin{table}[t!]
    \centering
    \renewcommand{\arraystretch}{1.2}
    \begin{tabular}{@{\hspace{0.8cm}}c@{\hspace{0.8cm}}c@{\hspace{0.6cm}}}
        \toprule[2.0pt]
        \toprule[1.0pt]
        Parameter & Value \\
        \midrule[1.0pt]
        $(\chi_1,\chi_2)$ & $(0,0)$ \\
        $\iota$ & $30^\circ$ \\
        $\psi$ & $0$ \\
        $\Phi_c$ & $0$ \\
        $t_c$ & $1126259642.413$ s \\
        $\alpha$  & $\pi/4$ rad \\
        $\delta$  & $0.85\cdot \pi/2$ rad \\
        $f_{\rm ref}$ & $100\,{\rm Hz}$ \\
        \bottomrule[1.0pt]
        \bottomrule[2.0pt]
    \end{tabular}
    \caption{
Parameters fixed for all injections considered in this work.
The total mass, mass ratio, and luminosity distance are varied among
the different analyses and are therefore not reported here.
}
\label{tab:fixed_injection_parameters}
\end{table}

The injections are generated by fixing the set of parameters reported in Table~\ref{tab:fixed_injection_parameters}, while varying the intrinsic binary properties and luminosity distance among the different analyses. In particular, the component tidal deformabilities are obtained from the adopted APR EoS relation $\Lambda(m)$~\cite{Akmal:1998cf} and are not treated as independent injection parameters. The binary tidal parameters $\tilde{\Lambda}$ and $\delta\tilde{\Lambda}$, as well as the higher-order tidal coefficients entering at 6PN, 6.5PN, 7PN, and 8PN, are subsequently computed from the component masses and deformabilities, as illustrated in Subsec.~\ref{subsec:TF2_wf}.

 For the injections, as performed in Ref.~\cite{Crescimbeni:2024cwh}, the taper width is fixed to
\begin{equation}
\Delta\lambda_f =
\frac{\lambda_f^{\rm inj}}{6},
\end{equation}
where $\lambda_f=f_{\rm cut}/f_{\rm ISCO}$. In the parameter estimation, instead, $\lambda_f$ is allowed to vary freely according to the prior reported in Appendix~\ref{app:priors}, along with the priors on the other parameters.

\subsection{The impact of 7PN terms}
\label{sec:7PN}

The first Bayesian analysis presented in this work investigates the measurability of the 7PN tidal contributions, both for GW170817-like and subsolar-mass BNSs.

All injections are generated using the complete tidal phase through 7PN order, including both the next-to-next-to-leading-order electric quadrupolar contribution and the leading electric octupolar contribution. We use three distinct recovery models in the analysis:
\begin{itemize}
    \item $\delta\psi_{\rm tidal}^{7{\rm PN}}$: complete 7PN tidal phase model~[see Eq.~\eqref{eq:A7PN}];

\item $\delta\psi_{\rm tidal}^{7{\rm PN}}|_{\Lambda_3=0}$: 7PN tidal phase model with the $\ell=3$ (octupolar) tidal contribution omitted;

\item $\delta \psi_{\rm tidal}^{6.5{\rm PN}}$: tidal phase model truncated at 6.5PN.
\end{itemize}

In all three cases, we determine the higher-order tidal coefficients using the adopted QURs, which allows us to separately assess the relevance of the tidal terms entering at 6.5 and 7PN. For instance, comparing the evidence of the complete 7PN recovery with that of the 6.5PN model quantifies the overall measurability of the 7PN contribution, while comparing the complete 7PN model with the 7PN model without the octupolar term isolates the impact of the leading octupolar correction.

We perform two complementary sets of analyses. In the first one, Table~\ref{tab:7PN_mass_summary}, we fix the detector SNR to 1800 and vary the source-frame total mass of the binary, while luminosity distance is adjusted for each configuration so as to keep the SNR fixed. This deliberately high-SNR benchmark allows us to isolate the dependence of the different 7PN contributions on the intrinsic binary parameters, without conflating it with variations in signal strength. Although such a high SNR is not representative of typical events, it is within the reach of 3G observatories for sufficiently nearby and favourably oriented sources, and may be further enhanced by combining multiple detectors in a network~\cite{galaxies10040090}.

In the second analysis, Table~\ref{tab:7PN_SNR_summary}, we fix the source-frame total mass and vary the luminosity distance, and hence the SNR. This complementary scan allows us to determine the SNR regime in which the complete 7PN contribution, and in particular the substantially smaller octupolar term, become distinguishable by Bayesian model selection.

\begin{table}[t!]
    \centering
    \renewcommand{\arraystretch}{1.2}
    \begin{tabular}{c@{\hspace{0.8cm}}c@{\hspace{0.8cm}}c@{\hspace{0.8cm}}c}
        \toprule[2.0pt]
        \toprule[1.0pt]
        $M_{\rm tot}\,[M_\odot]$ & $d_L\,[{\rm Mpc}]$ & $\log_{10}\mathcal{B}^{7{\rm PN}}_{6.5{\rm PN}}$ & $\log_{10}\mathcal{B}^{7{\rm PN}}_{\Lambda_3=0}$ \\
        \midrule[1.0pt]
        1.0 & 9 & $47.5$ & $3.4$ \\
        1.2 & 11 & $39.9$ & $1.9$ \\
        1.5 & 13 & $30.2$ & $0.9$ \\
        2.0 & 17 & $15.2$ & $0.1$ \\
        2.8 & 22 & $3.1$ & $-0.2$ \\
        \bottomrule[1.0pt]
        \bottomrule[2.0pt]
    \end{tabular}
    \caption{Bayesian model-selection results illustrating the mass dependence of the 7PN tidal corrections for binaries with $q=0.93$ at fixed $\text{SNR}=1800$.
    For each source-frame total mass $M_{\rm tot}$ we report the luminosity distance $d_L$ that yields this SNR and two Bayes factors of the complete 7PN recovery model, $\delta\psi_{\rm tidal}^{7{\rm PN}}$: $\mathcal{B}^{7{\rm PN}}_{6.5{\rm PN}}$ against the model truncated at 6.5PN, in which all 7PN tidal contributions are removed, and $\mathcal{B}^{7{\rm PN}}_{\Lambda_3=0}$ against the 7PN model in which only the electric-octupole contribution is omitted.
    Positive values of $\log_{10}\mathcal B$ favour the more complete waveform model.
    }
    \label{tab:7PN_mass_summary}
\end{table}

\begin{table}[t!]
    \centering
    \renewcommand{\arraystretch}{1.2}
    \begin{tabular}{@{\hspace{0.4cm}}c@{\hspace{0.8cm}}c@{\hspace{0.8cm}}c@{\hspace{0.8cm}}c@{\hspace{0.4cm}}}
        \toprule[2.0pt]
        \toprule[1.0pt]
        $\text{SNR}$ & $d_L\,[{\rm Mpc}]$ & $\log_{10}\mathcal{B}^{7{\rm PN}}_{6.5{\rm PN}}$ & $\log_{10}\mathcal{B}^{7{\rm PN}}_{\Lambda_3=0}$ \\
        \midrule[1.0pt]
        422 & 40 & $3.5$ & $0.7$ \\
        562 & 30 & $5.8$ & $0.6$ \\
        843 & 20 & $11.4$ & $0.8$ \\
        1687 & 10 & $41.8$ & $3.2$ \\
        \bottomrule[1.0pt]
        \bottomrule[2.0pt]
    \end{tabular}
    \caption{Same as Table~\ref{tab:7PN_mass_summary} but varying the luminosity distance (and hence the SNR) for subsolar binaries with fixed source-frame total mass $M_{\rm tot}=1\,M_\odot$ and $q=0.93$.
    }
    \label{tab:7PN_SNR_summary}
\end{table}

The two analyses reveal a clear hierarchy between the different contributions entering at 7PN order. At fixed $\text{SNR}=1800$, the complete 7PN contribution remains strongly favoured over the 6.5PN model across the entire mass range considered. The corresponding Bayes factor decreases with increasing total mass, from $\log_{10}\mathcal{B}\simeq48$ at $M_{\rm tot}=1\,M_\odot$ to $\log_{10}\mathcal{B}\simeq3.1$ at $M_{\rm tot}=2.8\,M_\odot$, always denoting a decisive measurability of the 7PN terms.

The octupolar contribution is considerably more difficult to resolve independently. For the same fixed SNR, omitting the octupolar term produces decisive evidence in favour of the complete model for $M_{\rm tot}=1\,M_\odot$, and strong evidence at $M_{\rm tot}=1.2\,M_\odot$. At $M_{\rm tot}=1.5\,M_\odot$ the evidence is only substantial and, for larger masses, the Bayes factor rapidly becomes inconclusive. This behaviour is consistent with the stronger tidal response of low-mass, less compact neutron stars and with the relative hierarchy of the 7PN quadrupolar and octupolar phase contributions discussed in Sec.~\ref{sec:waveform}.

On the other hand, for $M_{\rm tot}=1\,M_\odot$, the complete 7PN contribution is already decisively favoured at SNRs of a few hundred, and its Bayes factor grows rapidly as the SNR increases. In contrast, the octupolar term yields only substantial evidence for $\text{SNR}\lesssim850$, and becomes decisively distinguishable only for exceptionally loud signals with $\text{SNR}\gtrsim {\cal O}(10^3)$.

These results indicate that the complete 7PN tidal sector may become relevant well before we can separately measure all its individual physical contributions. Specifically, the next-to-next-to-leading quadrupolar correction dominates the measurability of the 7PN sector, while the octupolar contribution becomes observationally relevant primarily for exceptionally loud, low-mass binary neutron-star systems.

\subsection{The impact of 8PN terms}

We next investigate the measurability of the tidal contributions entering at 8PN order, following the same Bayesian model-selection procedure introduced above for the 7PN analysis.

\begin{figure}[!t]
    \centering
    \includegraphics[width=\columnwidth]{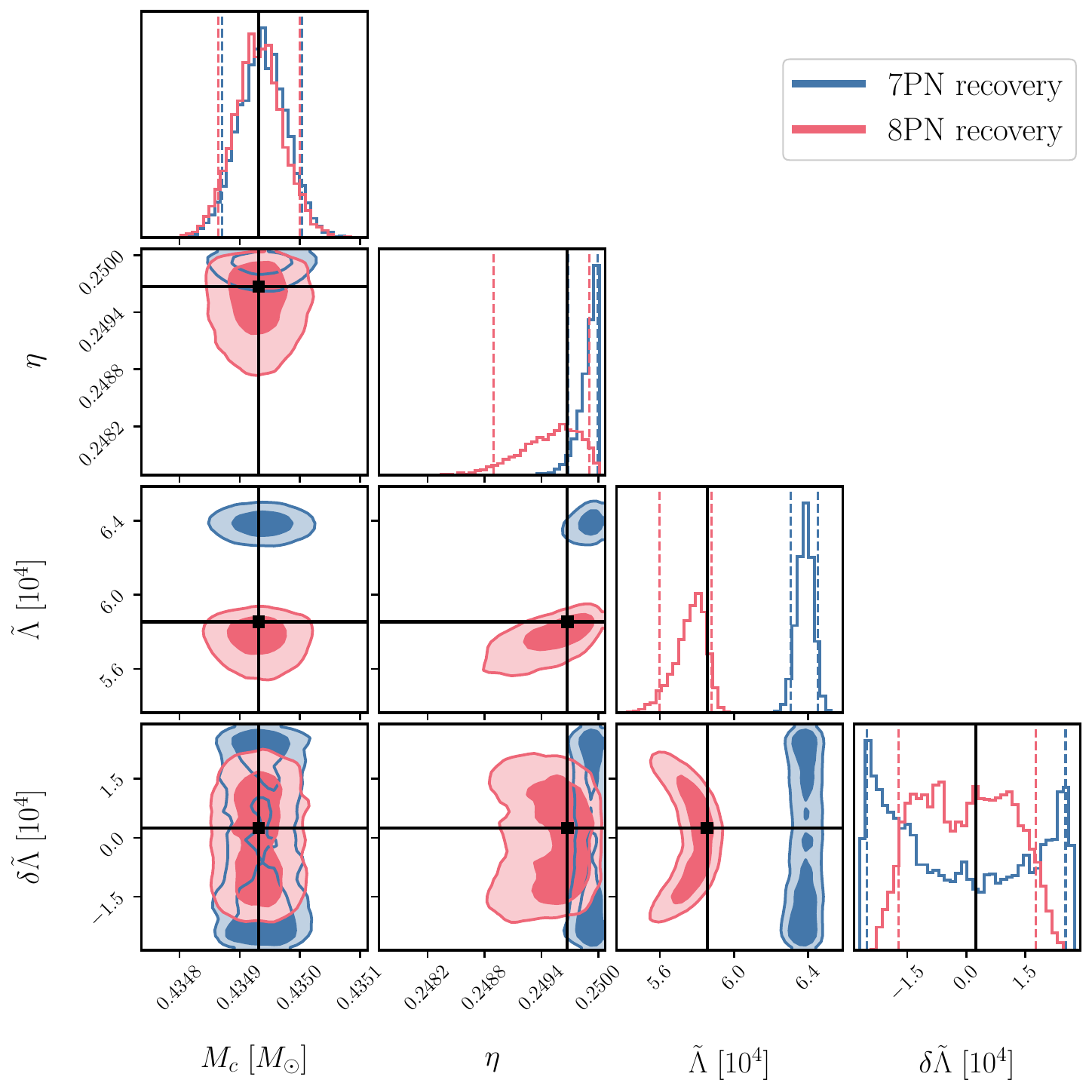}
    \caption{
    Posterior distributions for a binary with
    $M_{\rm tot}=1\,M_\odot$, $q=0.93$, and $\text{SNR}=200$,
    recovered with the complete QUR-based 8PN model (red) and with the
    QUR-based model truncated at 7PN (blue).
    The signal is injected with the full 8PN waveform.
    The inner and outer contours enclose $50\%$ and $90\%$ of the
    posterior probability, respectively, and the dashed lines in the
    one-dimensional panels bound the central $90\%$ credible intervals.
    Black lines denote the injected values;
    $M_c$ is the source-frame chirp mass.
    The full set of sampled parameters for the same injection is shown in Fig.~\ref{fig:corner_full_M1_snr200}.
    }
    \label{fig:corner_qur_M2p8}
\end{figure}

All injections are generated using the most complete waveform model considered in this work, which includes all tidal terms through 7PN supplemented by the leading-order dynamical and quadratic tidal contributions at 8PN [see Eq.~\eqref{eq:A8_total}]. Henceforth, we refer to this waveform as the \emph{full model}. We stress that it is not complete at 8PN: it omits the next-to-next-to-next-to-leading-order static electric-quadrupolar correction at 8PN, as well as other tidal contributions entering at 7.5PN and 8PN. We then perform the recovery using the following models:
\begin{itemize}
    \item $\delta\psi_{\rm tidal}^{\rm 7PN}$: complete 7PN tidal phase model;
    \item $\delta\psi_{\rm tidal}^{\rm 8PN}$: full model.
\end{itemize}
The comparison between the corresponding Bayesian evidences therefore quantifies the measurability of the overall 8PN tidal sector.

Unlike the 7PN analysis, where the octupolar contribution can be isolated by selectively removing it from the recovery waveform, here we initially consider the combined effect of the quadratic and dynamical 8PN corrections. We explore different binary masses and signal strengths to determine the region of parameter space in which the inclusion of the 8PN terms becomes statistically distinguishable from a waveform truncated at 7PN. An example is shown in Fig.~\ref{fig:corner_qur_M2p8}, which shows that injecting with $\delta\psi_{\rm tidal}^{8{\rm PN}}$, but recovering with $\delta\psi_{\rm tidal}^{7{\rm PN}}$ can induce systematic biases.

\begin{figure}[t]
    \centering
    \includegraphics[width=\columnwidth]{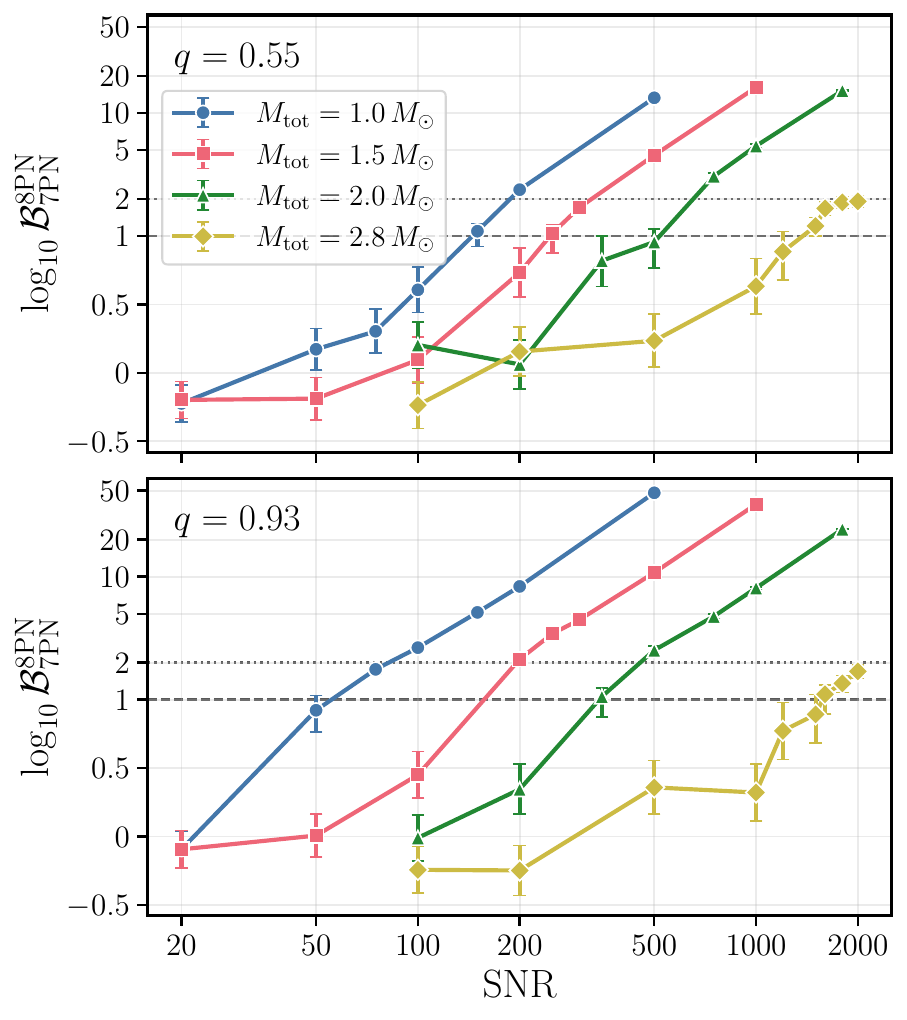}
    \caption{
    Bayes factor between the full 8PN recovery model and the waveform truncated at full 7PN order, shown as a function of the detector optimal SNR for binaries with different total masses $M_{\rm tot}$.
    The upper and lower panels correspond to mass ratios $q=0.55$ and $q=0.93$, respectively.
    The injected signal is generated with the complete waveform model including the 8PN tidal terms.
    The vertical axis shows
    $\log_{10}\mathcal{B}^{8{\rm PN}}_{7{\rm PN}}
    =
    \log_{10}(\mathcal Z_{\rm 8PN}/\mathcal Z_{\rm 7PN})$,
    with positive values favouring the inclusion of the 8PN tidal contributions; the axis is logarithmic above $\log_{10}\mathcal B = 1$ and linear below, where $\log_{10}B$ can become negative.
    The dashed and dotted horizontal lines mark $\log_{10}\mathcal B = 1$ and $\log_{10}\mathcal B = 2$, respectively, corresponding to reference thresholds in the Jeffreys scale.
    The error bars show the nested-sampling uncertainties on the two evidences, combined in quadrature.
    }
    \label{fig:bayes-factor-8PN-vs-7PN}
\end{figure}

We thus perform an SNR scan for four different source-frame total masses and for two representative mass ratios, $q=0.55$ and $q=0.93$. The corresponding results are shown in the upper and lower panels of Fig.~\ref{fig:bayes-factor-8PN-vs-7PN}, respectively. For each binary mass, the luminosity distance is varied to probe different SNRs, while keeping all other injected source properties fixed. Each point shown in the figure requires two independent Bayesian analyses. For each mass ratio, the four mass configurations are explored at seven to nine different SNRs.

As expected, the two mass-ratio scans show a clear dependence of the measurability of the 8PN tidal sector on both the total mass and the signal strength. For both values of the mass ratio, the Bayes factor grows rapidly with the SNR and decreases as the total mass of the system is increased.

Figure~\ref{fig:bayes-factor-8PN-vs-7PN} demonstrates that, for the configurations with $M_{\rm tot} \le 2 M_\odot$ considered here, the $8$PN model is favoured over the $7$PN model at lower SNR for nearly equal-mass BNSs than for more asymmetric BNSs of the same total mass. Specifically, in the more asymmetric case~$q=0.55$, the $8$PN model becomes clearly measurable at SNRs of $\sim150$ for $M_{\rm tot}=1\,M_\odot$, while strong evidence is reached only at SNRs of order $\sim250$ and $\sim500$ for $M_{\rm tot}=1.5\,M_\odot$ and $2\,M_\odot$, respectively. The $M_{\rm tot}=2.8\,M_\odot$ configuration reaches substantial evidence at $\text{SNR}\sim1000$ and strong evidence at $\text{SNR}\sim1500$.

By contrast, for nearly equal-mass binaries, $q=0.93$, the full 8PN contribution becomes strongly distinguishable already at $\text{SNR}\sim75$ for $M_{\rm tot}=1\,M_\odot$. As the total mass increases, louder signals are required: the transition to strong evidence occurs at SNRs of $\sim200$ for $M_{\rm tot}=1.5\,M_\odot$ and $\sim350$ for $M_{\rm tot}=2\,M_\odot$. For a canonical $M_{\rm tot}=2.8\,M_\odot$ binary, instead, the evidence becomes of order $\log_{10}\mathcal B\sim1$ only for $\text{SNR}\gtrsim1600$.

The impact of the 8PN sector is not limited to model selection. Fig.~\ref{fig:corner_qur_M2p8} shows the posterior distributions for a subsolar binary with $M_{\rm tot}=1\,M_\odot$ and $\text{SNR}=200$. The recovery truncated at 7PN overestimates $\tilde{\Lambda}$ by about $9\%$, with the injected values of both $\tilde{\Lambda}$ and $\eta$ lying outside their $90\%$ credible intervals, whereas the full 8PN recovery is consistent with the injection. Moreover, the 7PN recovery piles up at the edges of the range of $\delta\tilde{\Lambda}$ allowed by $\Lambda_2^{(1)},\Lambda_2^{(2)}>0$, whereas the 8PN recovery disfavors these edges, and its $90\%$ interval covers about $70\%$ of that range. A similar bias on $\tilde{\Lambda}$, of about $10\%$, is found for the canonical binary with $M_{\rm tot}=2.8\,M_\odot$ and $\text{SNR}=1800$, for which the Bayes factor is only of order $\log_{10}\mathcal{B}\sim1$.

\subsection{Agnostic measurement of the 8PN tidal contribution}
\label{sec:agnostic_8pn}

The analysis presented above relies on QURs to express the higher-order tidal coefficients in terms of the leading quadrupolar tidal deformabilities. While this provides a physically motivated reduction of the parameter space, it is also useful to assess how accurately the effective 8PN tidal contribution itself can be constrained without imposing these relations. To this end, we perform an additional set of parameter-estimation runs in which the total effective tidal coefficient entering at 8PN is treated as an independent parameter. For the injections, we retain both the dynamical and quadratic 8PN contributions and determine the corresponding higher-order tidal quantities consistently from the APR EoS and the adopted QURs through Eqs.~\eqref{eq:qur_quad} and~\eqref{eq:qur_dyn}. In the recovery, instead, we break the relation between the leading tidal parameters and the total 8PN coefficient, which we denote by $\Lambda_{8{\rm PN}}^{\rm tot}$ defined below.

For these \textit{agnostic} runs, we set the dynamical running coefficients $\beta_2^{(i)}$ to zero in the injections and assume its absence in the recovery. This removes the explicitly frequency-dependent logarithmic contribution from ${\cal A}_8$ given in Eq.~\eqref{eq:A8_total} and allows the full 8PN sector to be parametrized by a single effective, frequency-independent coefficient. We stress that this simplification is adopted only for the agnostic analysis, and the complete QUR-based analyses presented above retain the logarithmic running.

For the injections, this quantity is defined as
\begin{equation}
\begin{aligned}
 \Lambda_{8{\rm PN}}^{\rm tot}
    =&\sum_{i=1}^2X_i^5 \left(8X_i+147\eta\right)\hat{\Lambda}_{\rm dyn}^{(i)}+3\left(
        19\tilde{\zeta}_{+}
        -17\delta_{\eta}\,\tilde{\zeta}_{-}
    \right),
    \label{eq:lambda8_total}
\end{aligned}
\end{equation}
such that its contribution to the Fourier-domain phase is
\begin{equation}
    \mathcal{A}_{8}^{\rm agn}
    =
    -\frac{15}{11}
    \Lambda_{8{\rm PN}}^{\rm tot}.
    \label{eq:A8_agnostic}
\end{equation}
In the recovery, however, Eq.~\eqref{eq:lambda8_total} is no longer imposed, and $\Lambda_{8{\rm PN}}^{\rm tot}$ is sampled independently from $\tilde{\Lambda}$ and $\delta\tilde{\Lambda}$.

All the remaining phase terms of the waveform are kept identical to those used in the QUR-based analysis. In particular, the lower-order tidal sector and the tidal quantities entering at other PN orders are still determined through the adopted QURs.

For these exploratory runs, we adopt the broad uniform prior
\begin{equation}
    \Lambda_{8{\rm PN}}^{\rm tot}
    \sim
    {\cal U}\left(-5S_8,5S_8\right),
    \qquad
    S_8 =
    \max\left(
        \left|\Lambda_{8{\rm PN},{\rm inj}}^{\rm tot}\right|,
        1
    \right),
    \label{eq:lambda8_agnostic_prior}
\end{equation}
which allows both signs of the effective coefficient and places the injected value inside the prior range.

\begin{figure}[t]
    \centering
    \includegraphics[width=\columnwidth]{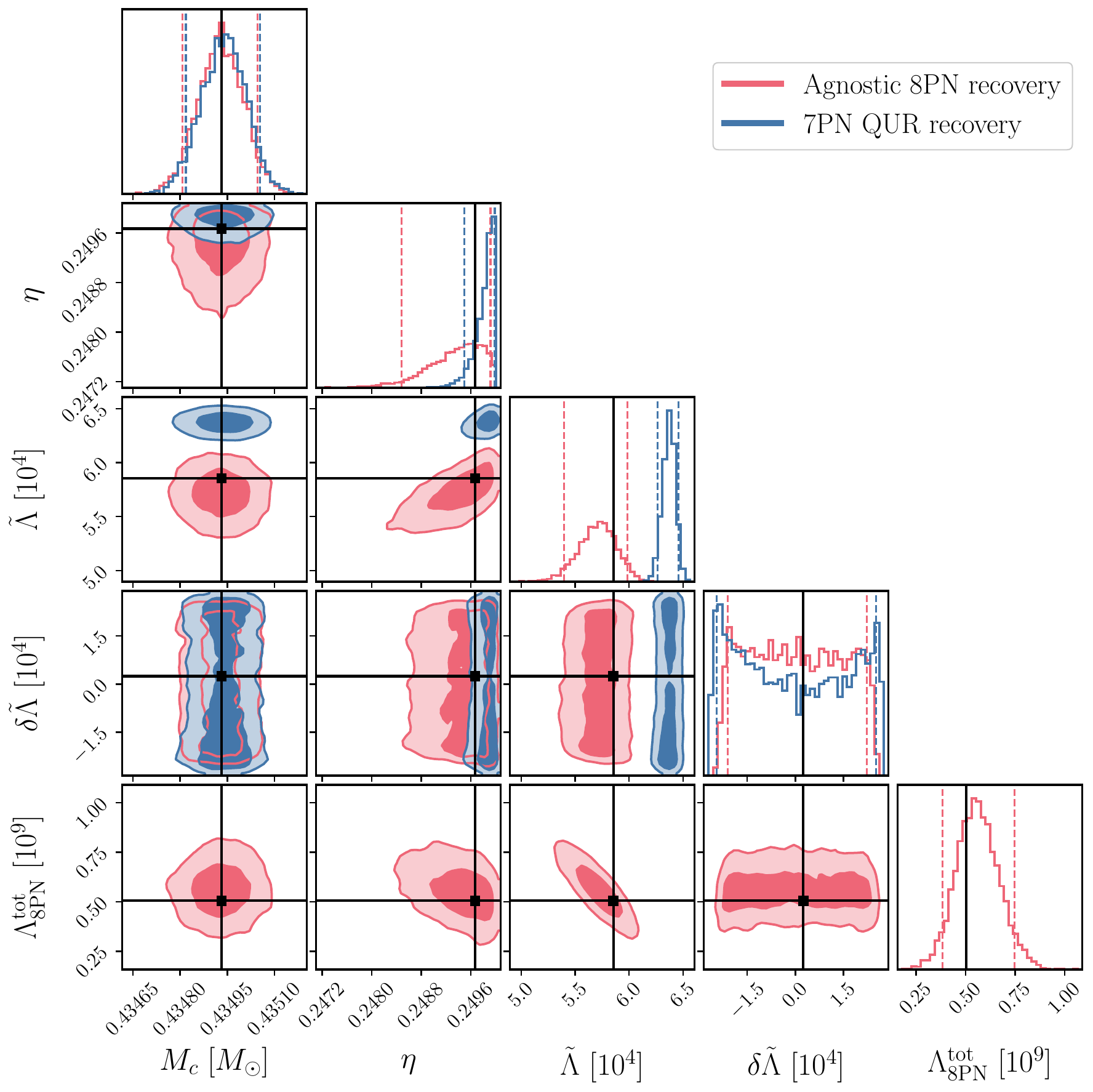}
    \caption{
    Posterior distributions for a binary with
    $M_{\rm tot}=1\,M_\odot$, $q=0.93$ and $\text{SNR}=150$,
    recovered with the agnostic 8PN model (red) and with the QUR-based
    model truncated at 7PN (blue).
    The signal is injected with the full 8PN waveform; in the
    agnostic recovery, the total 8PN tidal coefficient is sampled
    independently, while the 7PN model does not contain it, so the
    corresponding panels show the agnostic recovery only.
    The inner and outer contours enclose $50\%$ and $90\%$ of the
    posterior probability, respectively, and the dashed lines in the
    one-dimensional panels bound the central $90\%$ credible intervals.
    Black lines denote the injected values;
    $M_c$ is the source-frame chirp mass.
    }
    \label{fig:agnostic_vs_7PN_M1}
\end{figure}

\begin{figure}[t]
    \centering
    \includegraphics[width=\columnwidth]{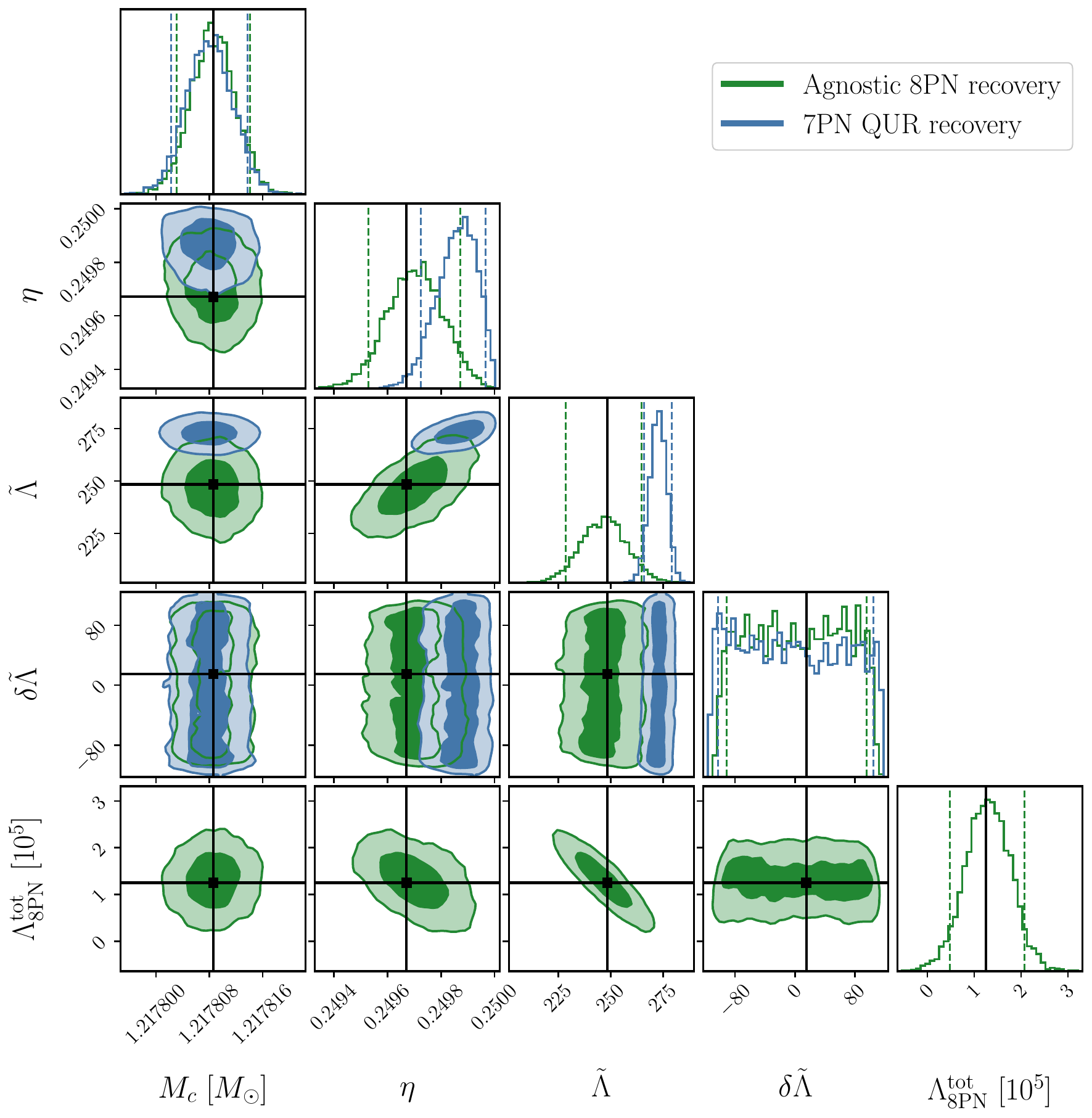}
    \caption{
    Same as Fig.~\ref{fig:agnostic_vs_7PN_M1}, but for $M_{\rm tot}=2.8\,M_\odot$, $q=0.93$, and $\text{SNR}=1800$, for the agnostic recovery (green) and the QUR-based recovery truncated at 7PN (blue), which does not contain $\Lambda_{8{\rm PN}}^{\rm tot}$. This case probes the high-SNR regime, where the effective total 8PN tidal contribution can be constrained despite being sampled
    independently in the recovery.
    }
    \label{fig:agnostic_corner_M2p8}
\end{figure}

In addition, the cutoff and tapering prescriptions are unchanged with respect to Sec.~\ref{subsec:frequencies}.
We consider representative configurations spanning the region in which the 8PN contribution becomes relevant according to the Bayes-factor analysis. Specifically, for $q=0.93$ we perform agnostic recoveries at $M_{\rm tot}=1\,M_\odot$ with SNRs $150$ and $300$, at $M_{\rm tot}=1.5\,M_\odot$ with SNRs $400$ and $800$, and at $M_{\rm tot}=2.8\,M_\odot$ with SNR $1800$. Representative posterior distributions are shown in Figs.~\ref{fig:agnostic_vs_7PN_M1} and~\ref{fig:agnostic_corner_M2p8}, for $M_{\rm tot}=1\,M_\odot$ at $\text{SNR}=150$ and for $M_{\rm tot}=2.8\,M_\odot$ at $\text{SNR}=1800$; the other configurations are qualitatively the same and are not shown.

The agnostic posterior distributions show that the effective 8PN tidal coefficient can be constrained independently of the lower-order tidal parameters. At fixed total mass, the measurement improves significantly as the SNR increases: for $M_{\rm tot}=1\,M_\odot$, the $90\%$ interval of $\Lambda_{8{\rm PN}}^{\rm tot}$ halves between $\text{SNR}=150$ and $300$. The posteriors also present a correlation between $\tilde{\Lambda}$ and $\Lambda_{8{\rm PN}}^{\rm tot}$, reflecting the partial degeneracy between the leading and higher-order tidal contributions to the phase. By contrast, $\delta\tilde{\Lambda}$ remains only weakly constrained in all the (almost equal-mass) configurations considered.

Figure~\ref{fig:agnostic_vs_7PN_M1} compares the agnostic recovery with a recovery of the same signal truncated at 7PN. Without the 8PN term, $\tilde{\Lambda}$ is overestimated by about $9\%$, with the injected value lying outside its $90\%$ interval, whereas $\eta$ and the chirp mass remain consistent with the injection. In the agnostic recovery, instead, $\tilde{\Lambda}$ is consistent with the injection, at the price of a $90\%$ interval of $\tilde{\Lambda}$ about three times as wide.

The 8PN contribution remains measurable even in the most challenging configuration considered here, $M_{\rm tot}=2.8\,M_\odot$, for which both stars have masses above $1\,M_\odot$ and tidal deformabilities substantially smaller than in the subsolar-mass cases. For a sufficiently loud event, with $\text{SNR}=1800$, the posterior of $\Lambda_{8{\rm PN}}^{\rm tot}$ is still localized around its injected nonzero value (see Fig.~\ref{fig:agnostic_corner_M2p8}). This is facilitated by the nearly equal-mass configuration: for $q=0.93$, $\delta_\eta\simeq0.04$, so the term proportional to $\delta\tilde{\Lambda}$ in $\mathcal{A}_6$ [Eq.~\eqref{eq:A6}] is strongly suppressed, and the poorly constrained $\delta\tilde{\Lambda}$ does not significantly degrade the measurement of $\Lambda_{8{\rm PN}}^{\rm tot}$. As for $M_{\rm tot}=1\,M_\odot$, the recovery truncated at 7PN, also shown in Fig.~\ref{fig:agnostic_corner_M2p8}, overestimates $\tilde{\Lambda}$, here by about $10\%$, and excludes the injected values of both $\tilde{\Lambda}$ and $\eta$ from its $90\%$ credible intervals. Finally, compared with the QUR-based 8PN recovery of the same signal (not shown), imposing the QURs narrows the $90\%$ interval of $\tilde{\Lambda}$ by about $60\%$.

As a complementary test, we compare the Bayesian evidence obtained with the agnostic 8PN parametrization against that of the corresponding QUR-based recovery. We consider the two representative injection configurations for which matching QUR runs are available, for $M_{\rm tot}=1\,M_{\odot}$ and $M_{\rm tot}=2.8M_{\odot}$, where the SNRs are respectively 150 and 1800. In both cases the QUR-based model is moderately preferred over the agnostic parametrization, with $\log_{10}\mathcal{B}_{\rm QUR/Agn}=0.85$ and $0.91$, respectively. Thus, while the data allow the total 8PN coefficient to be constrained independently, they remain compatible with the more restrictive QUR-based description.

\subsection{Mass--deformability reconstruction}

We reconstruct the source-frame component masses and the individual
quadrupolar tidal deformabilities from each posterior sample, obtaining
$\Lambda_2^{(1)}$ and $\Lambda_2^{(2)}$ from $\tilde{\Lambda}$ and
$\delta\tilde{\Lambda}$ through Eqs.~\eqref{LT} and~\eqref{dLT}. We compare the
resulting posterior regions in the $(m,\Lambda)$ plane with theoretical
mass--deformability relations to assess how the treatment of the 8PN
tidal sector affects the inferred stellar properties and their
implications for the EoS.

\begin{figure}[t]
    \centering
    \begin{minipage}{0.49\textwidth}
        \centering
        \includegraphics[width=\textwidth]{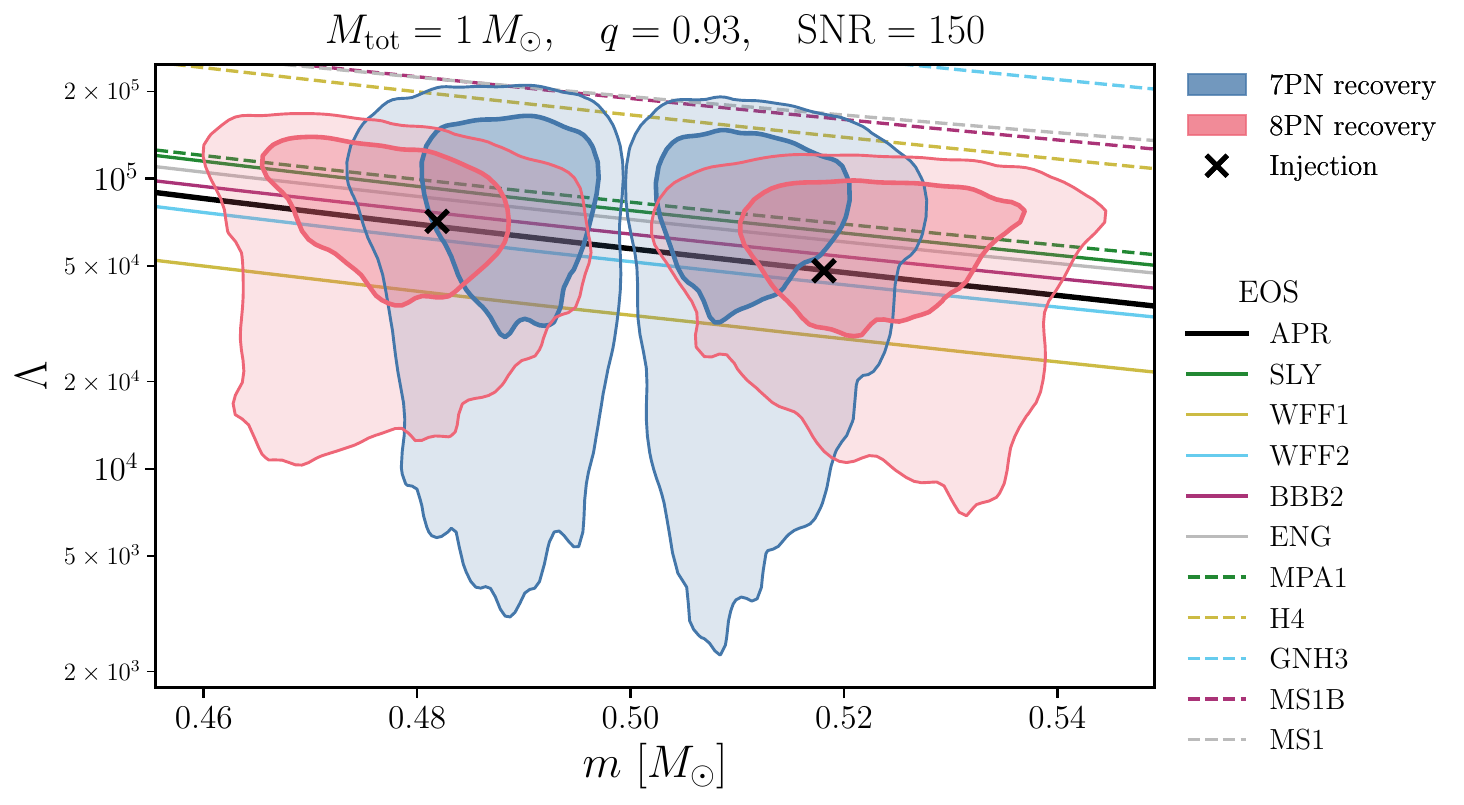}
        \textbf{(a)}
    \end{minipage}
    \hfill
    \begin{minipage}{0.49\textwidth}
        \centering
        \includegraphics[width=\textwidth]{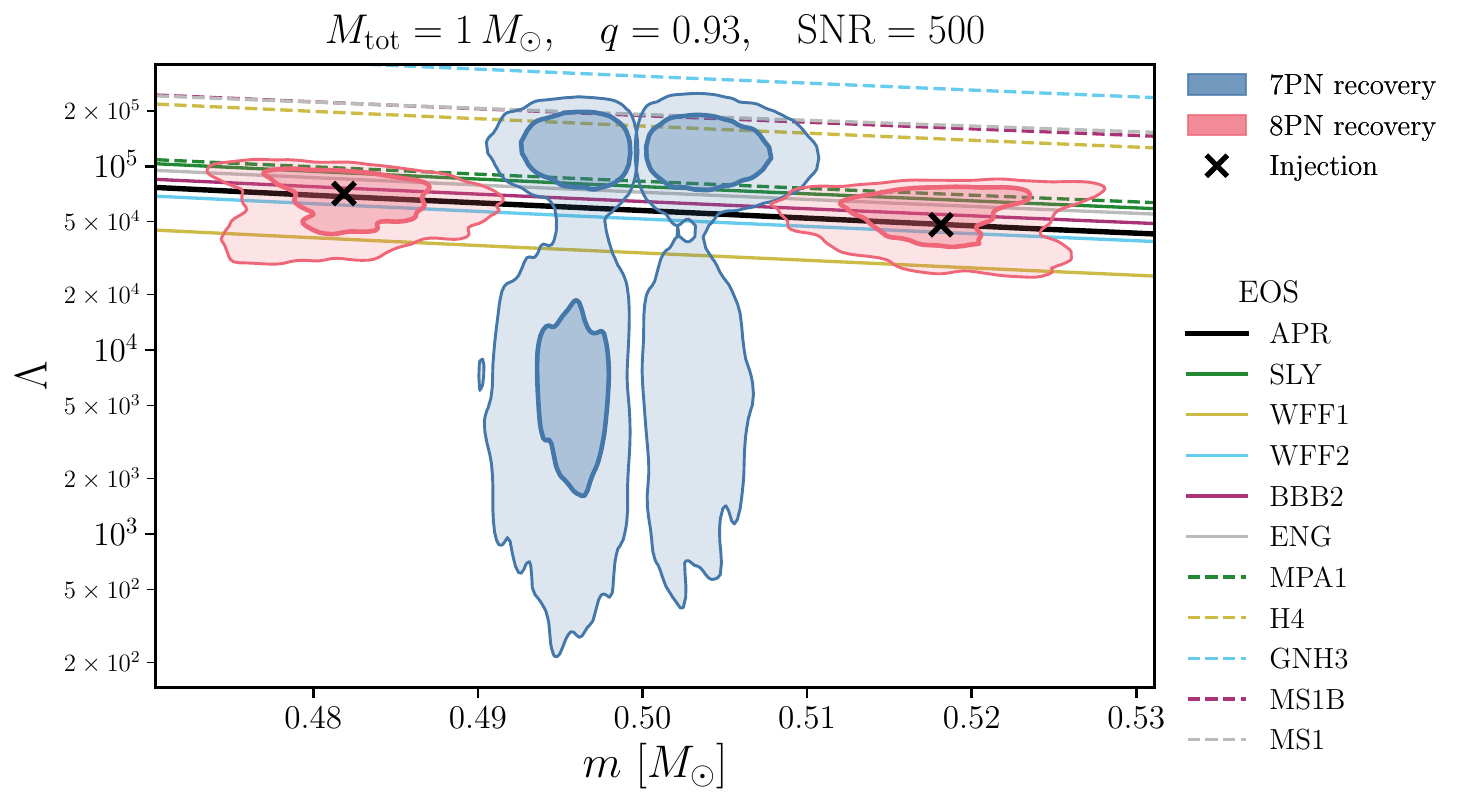}
        \textbf{(b)}
    \end{minipage}
    \caption{
        Posterior constraints in the component-mass--tidal-deformability
        plane from the QUR-based analyses of a binary with
        $M_{\rm tot}=1\,M_\odot$ and $q=0.93$.
        The injected waveform includes tidal effects through 8PN.
        Blue regions correspond to a recovery truncated at 7PN,
        whereas red regions correspond to the full 8PN recovery.
        The inner and outer contours enclose $50\%$ and $90\%$ of the
        posterior probability, respectively.
        Black crosses denote the injected component masses and
        quadrupolar tidal deformabilities.
        The curves show the mass--deformability relations predicted
        by representative neutron-star EoSs.
        Panels (a) and (b) correspond to $\text{SNR}=150$ and
        $500$, respectively.
    }
    \label{fig:eos_comparison_M1}
\end{figure}

We first consider the QUR-based analyses.
Figure~\ref{fig:eos_comparison_M1} shows the reconstruction for
$M_{\rm tot}=1\,M_\odot$ and $q=0.93$, using an injected signal that
includes tidal effects through 8PN. At $\text{SNR}=150$, both
recoveries overlap several of the EoS curves shown, but the complete
8PN recovery is centered closer to the injected masses and
deformabilities. At $\text{SNR}=500$, the posterior regions become
narrower and the difference between the two recoveries is more apparent:
the 7PN recovery retains a different correlation structure in the
$(m,\Lambda)$ plane, whereas the complete recovery is localized around
the injected values. For this system, including the 8PN terms primarily
improves the reconstruction of the stellar properties, without a
comparably pronounced improvement in the discrimination among the
EoS shown.

\begin{figure}[t]
    \centering
    \begin{minipage}{0.49\textwidth}
        \centering
        \includegraphics[width=\textwidth]{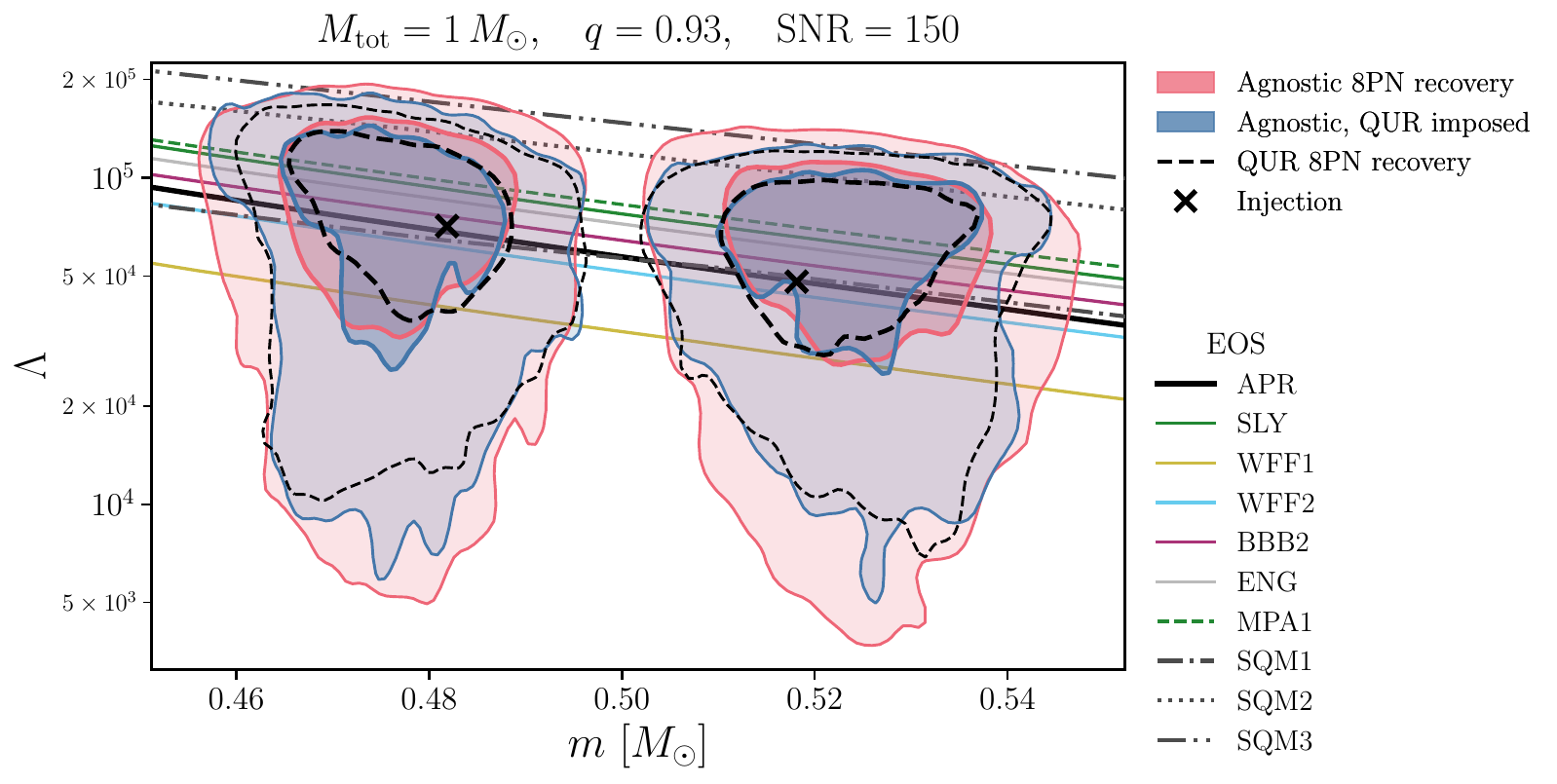}
        \textbf{(a)}
    \end{minipage}
    \hfill
    \begin{minipage}{0.49\textwidth}
        \centering
        \includegraphics[width=\textwidth]{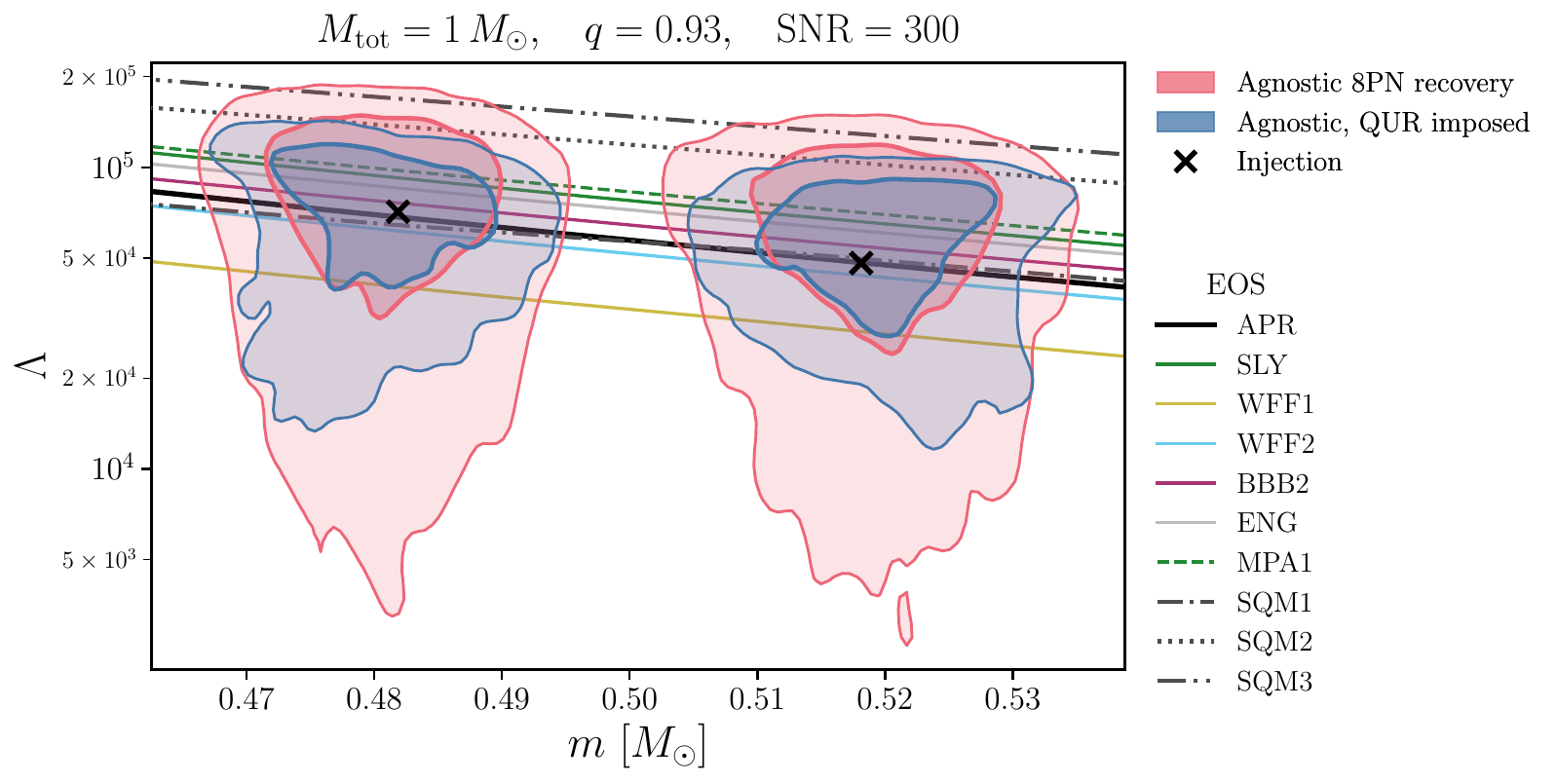}
        \textbf{(b)}
    \end{minipage}
    \caption{
        Posterior constraints in the component-mass--tidal-deformability
        plane obtained with the agnostic 8PN recovery for
        $M_{\rm tot}=1\,M_\odot$ and $q=0.93$.
        Red regions correspond to the agnostic recovery, whereas
        blue regions are obtained by reweighting the same posterior
        samples with the QUR-consistency weight of
        Eq.~\eqref{eq:qur_reweighting}, using
        $\sigma_\Delta=0.07\,S_8$.
        The inner and outer contours enclose $50\%$ and $90\%$ of
        the posterior probability, respectively, and black crosses
        denote the injected values.
        Dashed black contours in panel (a) show the direct
        QUR-based 8PN recovery.
        Coloured curves show representative ordinary neutron-star EoSs, while dark grey dash-dotted and dotted curves
        show the SQM1, SQM2, and SQM3 strange-quark-matter models
        considered in Ref.~\cite{Crescimbeni:2024qrq}.
        Panels (a) and (b) correspond to $\text{SNR}=150$ and
        $300$, respectively.
    }
    \label{fig:eos_agnostic_M1}
\end{figure}

We next consider the agnostic 8PN analysis, in which
$\Lambda_{8{\rm PN}}^{\rm tot}$ is sampled independently of the
leading tidal parameters. The red regions in
Fig.~\ref{fig:eos_agnostic_M1} show the resulting reconstruction for
$M_{\rm tot}=1\,M_\odot$ and $q=0.93$. Increasing
$\text{SNR}$ from $150$ to $300$ substantially tightens the mass
constraints, while the posterior regions remain close to the injected
values. The individual deformabilities are less tightly constrained:
even at $\text{SNR}=300$, their $90\%$ regions extend to values
several times smaller than those injected.

The agnostic analysis also allows us to test whether the independently
inferred 8PN coefficient is consistent with the QUR prediction.
For each posterior sample, we compute
$\Lambda_{8{\rm PN}}^{\rm QUR}$ from
Eq.~\eqref{eq:lambda8_total}, using the reconstructed component
parameters and the QURs for the quadratic and dynamical Love numbers.
We set $\beta_2^{(i)}=0$, consistently with the agnostic waveform model,
and define
\begin{equation}
    \Delta \equiv
    \Lambda_{8{\rm PN}}^{\rm tot}
    - \Lambda_{8{\rm PN}}^{\rm QUR}.
    \label{eq:delta_qur}
\end{equation}
The value $\Delta=0$ lies within the $90\%$ credible interval in all
five configurations considered. The posterior medians range from
$-0.1\,S_8$ to $0$, where $S_8$ is the scale defined in
Eq.~\eqref{eq:lambda8_agnostic_prior}. The agnostic measurements are
therefore compatible with the QUR prediction at this credibility level.

We then investigate how requiring consistency with the QURs changes
the reconstruction. We assign each agnostic posterior sample the weight
\begin{equation}
    w(\Delta) \propto
    \exp\left[-\frac{\Delta^2}{2\sigma_\Delta^2}\right].
    \label{eq:qur_reweighting}
\end{equation}
The width $\sigma_\Delta$ specifies the allowed departure from the
QUR prediction: samples for which the independently inferred and
QUR-predicted 8PN coefficients agree retain substantial weight,
whereas samples with discrepancies much larger than $\sigma_\Delta$
are suppressed. This procedure adds a QUR-consistency constraint to
the agnostic posterior. The width is an adopted tolerance, rather
than a measured uncertainty, and a finite value does not impose the
exact condition $\Delta=0$.

The blue regions in Fig.~\ref{fig:eos_agnostic_M1} are obtained with
$\sigma_\Delta=0.07\,S_8$. Reweighting strongly suppresses the tails
towards small deformabilities: the lower endpoint of the $90\%$
credible interval for $\Lambda_2^{(1)}$ increases by a factor of
approximately $1.8$ at $\text{SNR}=150$ and $3.3$ at
$\text{SNR}=300$. The width of the corresponding interval for
$\tilde{\Lambda}$ decreases by about $40\%$ and $20\%$, respectively.
For comparison, the dashed black contours in panel (a) show the
direct QUR-based 8PN recovery.

This effect arises because the independently inferred~$\Lambda_{8{\rm PN}}^{\rm tot}$, together with the QURs, imposes an additional constraint on the individual static electric quadrupolar deformabilities~$\Lambda_2^{(1)}$ and $\Lambda_2^{(2)}$, in addition to that provided by $\tilde{\Lambda}$, thereby partially breaking the degeneracy between them. Requiring the QUR-predicted $8$PN coefficient to agree with the inferred value therefore
suppresses some of the asymmetric combinations that populate the
low-deformability tails. Near the minimum of the QUR prediction,
however, the 8PN coefficient varies weakly with the relative
deformabilities of the individual stars, leaving a residual degeneracy between them.

This effect persists when the tolerance is relaxed.
Using $\sigma_\Delta=0.3\,S_8$ still raises the lower endpoint of
the $90\%$ interval for $\Lambda_2^{(1)}$ by factors of approximately $1.3$ and $1.7$ at $\text{SNR}=150$ and $300$,
while reducing $\sigma_\Delta$ below $0.1\,S_8$ produces only a moderate
further change. This checks the sensitivity of the reconstruction
to the adopted tolerance, without assuming that this tolerance
represents a calibrated intrinsic scatter of the QURs.
For $M_{\rm tot}=2.8\,M_\odot$, where the fractional uncertainty
on $\Lambda_{8{\rm PN}}^{\rm tot}$ is approximately $40\%$, the same
reweighting reduces the width of the $\tilde{\Lambda}$ interval
by about $60\%$ but does not appreciably suppress the
low-deformability tail. A tighter constraint on the combined tidal
parameter therefore need not translate into tighter lower bounds
on the individual deformabilities.

\begin{figure}[t]
    \centering
    \begin{minipage}{0.49\textwidth}
        \centering
        \includegraphics[width=\textwidth]{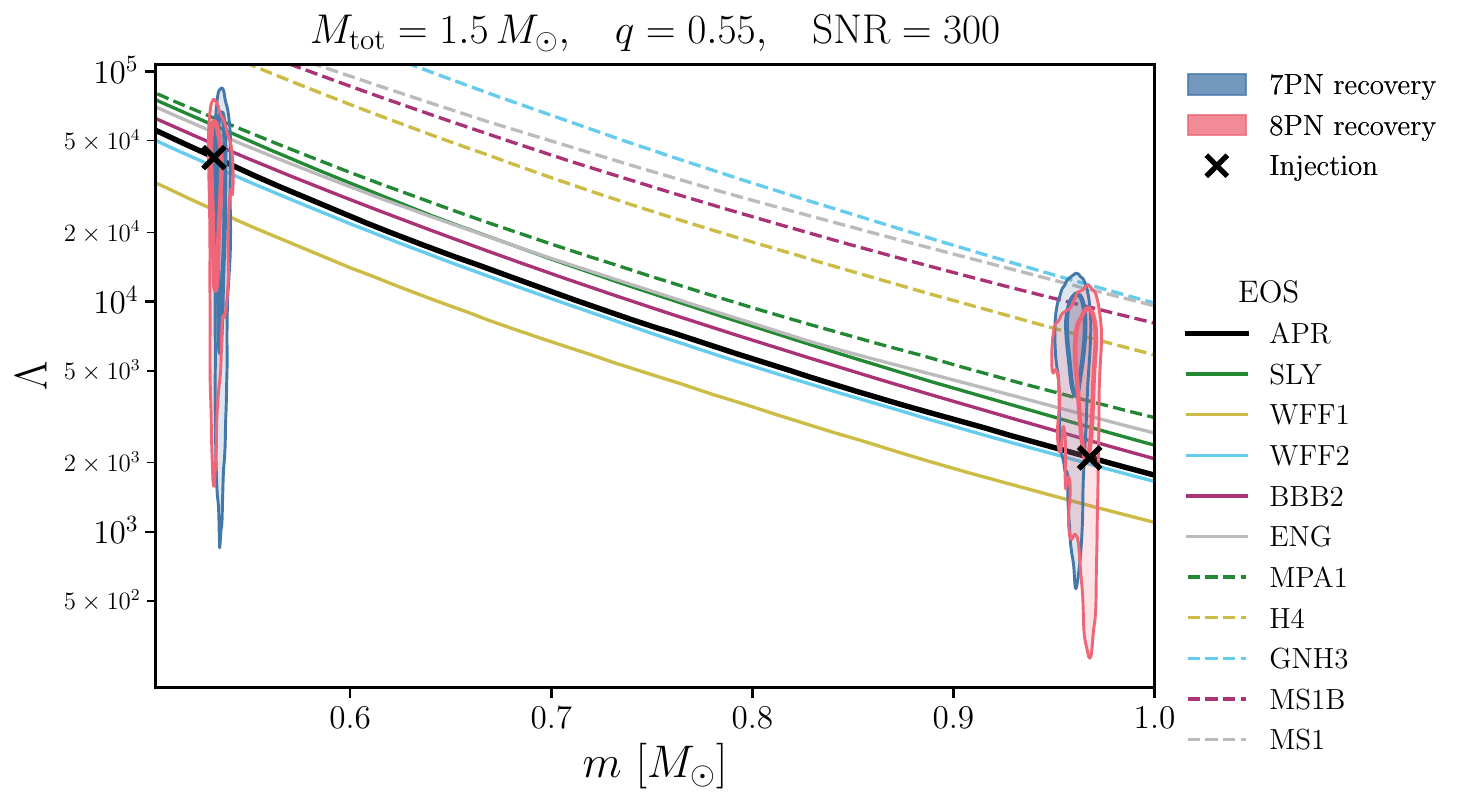}
        \textbf{(a)}
    \end{minipage}
    \hfill
    \begin{minipage}{0.49\textwidth}
        \centering
        \includegraphics[width=\textwidth]{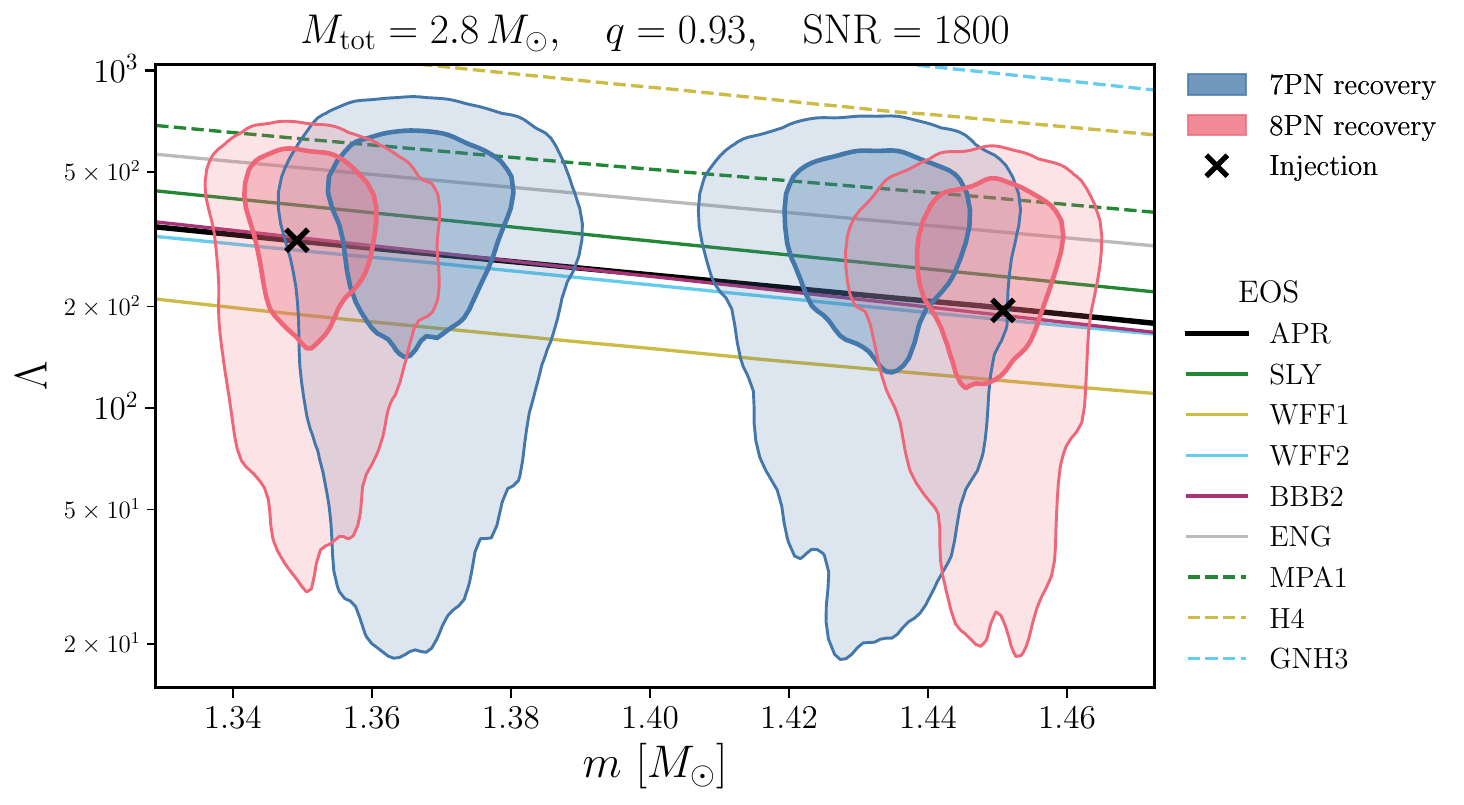}
        \textbf{(b)}
    \end{minipage}
    \caption{
        Posterior constraints in the component-mass--tidal-deformability
        plane obtained with the QUR-based analyses.
        Blue regions correspond to a recovery truncated at 7PN,
        whereas red regions correspond to the full 8PN recovery.
        The inner and outer contours enclose $50\%$ and $90\%$ of
        the posterior probability, respectively, and black crosses
        denote the injected values.
        The curves show the mass--deformability relations predicted
        by representative neutron-star EoSs.
        Panel (a) corresponds to $M_{\rm tot}=1.5\,M_\odot$,
        $q=0.55$, and $\text{SNR}=300$.
        Panel (b) corresponds to $M_{\rm tot}=2.8\,M_\odot$,
        $q=0.93$, and $\text{SNR}=1800$.
    }
    \label{fig:eos_q055_M1_and_M2p8}
\end{figure}

To place these results in the context of the other binary
configurations, Fig.~\ref{fig:eos_q055_M1_and_M2p8} shows the
QUR-based reconstruction for an unequal-mass system with
$M_{\rm tot}=1.5\,M_\odot$, $q=0.55$, and $\text{SNR}=300$,
and for a system in the canonical neutron-star mass range with
$M_{\rm tot}=2.8\,M_\odot$, $q=0.93$, and $\text{SNR}=1800$.
In the unequal-mass case, the two components probe well-separated
portions of the mass--deformability relations, but their
deformabilities remain broadly distributed under both recoveries.
In the higher-mass case, the 7PN and 8PN posterior regions overlap
substantially, and extended tails towards small deformabilities
persist even at this high SNR. The effect of the 8PN terms thus
depends on the binary configuration and does not produce a uniform
reduction of the uncertainties in the individual deformabilities.

The EoS curves in these figures illustrate the stellar-structure
information contained in the reconstructed posteriors.
In Fig.~\ref{fig:eos_agnostic_M1}, we also include the SQM1, SQM2,
and SQM3 strange-quark-matter sequences considered in
Ref.~\cite{Crescimbeni:2024qrq}. Sequences close to the injected
APR relation remain difficult to distinguish, whereas more distant
sequences have less overlap with the tightly constrained posterior
regions. This visual comparison does not assign probabilities to
the EoS models. A quantitative comparison would require imposing
each EoS-dependent mass--deformability relation consistently for
both components and accounting for the corresponding priors.

\section{Discussion and Conclusions}
\label{sec:conclusions}
In this work, we investigated the observational impact of higher-order tidal corrections in BNS inspirals, focusing on the leading electric-octupolar contribution at 7PN order and the recently derived dynamical and quadratic tidal effects at 8PN order. We incorporated these corrections into a frequency-domain waveform approximant, implemented in the publicly available Python package \texttt{TaylorTidal8PN}~\cite{Valletta:TaylorTidal8PN}, and performed Bayesian analyses for binaries spanning the canonical neutron-star and subsolar-mass regimes. Our results show that higher-order tidal effects can become measurable with next-generation GW detectors and, if neglected, may introduce significant systematic biases in the inferred neutron-star properties.

At 7PN order, we find a clear hierarchy between the quadrupolar and octupolar contributions. The complete 7PN sector can be distinguished from a waveform truncated at 6.5PN order at detector SNRs of a few hundred for low-mass binaries, with the measurability dominated by the next-to-next-to-leading-order quadrupolar correction. In contrast, isolating the leading electric-octupolar contribution requires considerably louder signals, with SNRs of order $10^3$ even for nearly equal-mass subsolar binaries. Thus, the complete 7PN tidal sector becomes observationally relevant well before its individual contributions can be independently resolved.

The 8PN tidal sector, comprising the dynamical and quadratic tidal responses, displays an even stronger dependence on the binary masses. For nearly equal-mass binaries with $M_{\rm tot}=1\,M_\odot$, the inclusion of the full 8PN contribution is strongly favoured already at detector SNRs of order $100$. The corresponding threshold increases with total mass, reaching SNRs of order $10^3$ for canonical BNS systems with $M_{\rm tot}=2.8\,M_\odot$. This behaviour reflects the enhancement of higher-order tidal effects at low compactness, which can partially compensate for their nominal PN suppression. These results suggest that the 8PN corrections may be particularly important for future observations of putative subsolar-mass neutron stars, while exceptionally loud events will be required to probe them in binaries with canonical component masses.

Interestingly, the measurability of the 8PN sector does not simply follow the magnitude of its contribution to the GW phase. Although individual 8PN coefficients can be enhanced in asymmetric binaries, our Bayesian analyses generally find stronger evidence for the full 8PN waveform in nearly equal-mass systems at fixed total mass and SNR. This illustrates the important role of correlations among tidal parameters: the distinguishability of higher-order corrections depends not only on their absolute magnitude, but also on the extent to which their effects can be mimicked by variations of the lower-order parameters. In particular, the combination $\delta\tilde{\Lambda}$ is poorly constrained for nearly equal-mass binaries, whereas its contribution and correlations become more relevant as the mass asymmetry increases. The interplay between these degeneracies and the additional frequency dependence introduced by the 8PN terms affects the resulting Bayesian evidence. Therefore, estimates based solely on the relative magnitude of the PN coefficients may not reliably predict the observational relevance of individual tidal corrections.

We complemented the analysis based on approximately EoS-insensitive relations with an agnostic approach, in which the effective 8PN tidal coefficient $\Lambda_{\rm 8PN}^{\rm tot}$, which combines the dynamical and quadratic tidal deformabilities, is treated as an independent parameter. In this case, we neglect the logarithmic running of the dynamical tidal response, allowing the combined 8PN contribution to be described by a single frequency-independent coefficient. We find that this coefficient can be constrained independently of the leading tidal deformabilities, even for canonical-mass binaries, provided the SNR is sufficiently large. The resulting posteriors exhibit a correlation between $\Lambda_{\rm 8PN}^{\rm tot}$ and $\tilde{\Lambda}$, reflecting the partial degeneracy between the leading- and higher-order tidal responses. Nevertheless, the inferred 8PN coefficient remains compatible with the QUR prediction in all the configurations considered. For the two configurations in which we directly compared the agnostic and QUR-based models, the latter is moderately preferred, indicating that the additional freedom introduced by the agnostic parametrization is not required by the simulated signals.

An important consequence of these results concerns the reconstruction of neutron-star properties. Recovering signals containing the full 8PN tidal sector with a waveform truncated at 7PN can produce appreciable biases in the inferred masses and tidal deformabilities. For instance, in the subsolar-mass configuration with $M_{\rm tot}=1\,M_\odot$ and detector SNR $150$, neglecting the 8PN contribution leads to an overestimation of $\tilde{\Lambda}$ by approximately $9\%$, with the injected value lying outside the $90\%$ credible interval. Including the 8PN corrections restores consistency with the injected parameters, although allowing the effective 8PN coefficient to vary independently generally increases the statistical uncertainties on the leading tidal deformabilities. This illustrates the trade-off between the systematic errors introduced by incomplete waveform models and the statistical uncertainties associated with additional tidal parameters.

The reconstruction of the individual stellar deformabilities reveals further subtleties. Although the effective 8PN coefficient can be independently measured, this does not necessarily translate into precise constraints on the tidal deformabilities of the two components. In particular, extended posterior tails towards small deformabilities can persist even at high SNR. Nevertheless, the leading- and higher-order tidal coefficients probe different combinations of the leading-order stellar responses.
Specifically, imposing consistency between the agnostically inferred effective $8$PN coefficient and the QUR predictions can partially break the degeneracies between the individual leading-order tidal deformabilities, thereby improving their reconstruction. Our comparison with representative mass-deformability relations shows that these improvements can be relevant for EoS inference, although distinguishing between models predicting similar tidal properties remains challenging. A quantitative assessment of EoS discrimination would require a dedicated model-selection analysis incorporating the corresponding mass--deformability relations and priors.

Our analysis relies on several simplifying assumptions. We adopt a PN inspiral waveform with a physically motivated cutoff frequency, estimated from the onset of tidal disruption or the Schwarzschild ISCO, to avoid extrapolating the high-order tidal corrections into the merger regime.
Furthermore, we considered a single ET observatory in its triangular configuration and fixed the source sky position and inclination to their injected values. This choice minimizes parameter correlations and allows us to isolate the impact of individual tidal contributions, which is the primary goal of this work, while reducing potential degeneracies that could obscure the effects of higher-order tidal terms.
Effects associated with the finite detector size, Earth's rotation, and extending the analysis to lower frequencies are expected to affect primarily the low-frequency signal and the inference of masses and spins~\cite{Crescimbeni:2026dvg}, although their interplay with the tidal parameters deserves further investigation. The quantitative results also depend on the adopted EoS and QURs, and a systematic exploration of these dependencies is left for future work.

Several extensions are particularly relevant. A systematic study of additional tidal contributions, such as rotational tidal Love numbers~\cite{JimenezForteza:2018rwr,Pani:2018inf,Gupta:2023oyy}, would allow a more complete assessment of finite-size effects, particularly for spinning neutron stars.
Furthermore, given the dominant contribution of the subleading quadrupole term at 7PN, it would be important to compute its contribution at the next order, i.e. include it at 8PN and compare it with the dynamical and quadratic tidal terms considered here.

It will also be important to go beyond the low-frequency expansion of the dynamical tidal response considered here and investigate the excitation of stellar oscillation modes, especially as the binary approaches merger~\cite{Gupta:2023oyy,Gittins:2026ntx}. More generally, incorporating the higher-order tidal corrections into effective-one-body waveform models would provide a natural framework to extend the present analysis into the strong-field regime and assess the robustness of our conclusions against more complete inspiral-merger descriptions (see, e.g., Ref.~\cite{Schulze:2026ewu}).

Overall, our results demonstrate that higher-order tidal corrections, including the recently derived dynamical and quadratic Love-number contributions at 8PN order, can play an important role in precision GW observations of neutron-star binaries. Their impact is particularly pronounced in the subsolar-mass regime, where the enhanced tidal response can make formally high-PN-order effects observationally relevant. Accurate modelling of these contributions will therefore be important not only to avoid systematic biases in future high-SNR observations, but also to exploit their potential for probing the nonlinear and dynamical response of neutron-star matter.

\begin{acknowledgments}
We thank Simone D'Onofrio for useful discussions on the draft. Inference analyses have been performed at the Vera and CHRONOS clusters supported by MUR and Sapienza University of Rome and at CINECA. We acknowledge financial support by the MUR FIS2 Advanced Grant ET-NOW (CUP:~B53C25001080001) and by the INFN TEONGRAV initiative.
\end{acknowledgments}

\appendix

\section{Sampling priors}
\label{app:priors}

The parameter estimation analyses employ broad priors on the sampled parameters shown in Table~\ref{tab:priors}. The additional prior associated with the effective 8PN coefficient in the agnostic analyses is discussed separately in
Sec.~\ref{sec:agnostic_8pn}.

\begin{table}[t!]
    \centering
    \renewcommand{\arraystretch}{1.2}
    \begin{tabular}{@{\hspace{0.5cm}}c@{\hspace{1.0cm}}c@{\hspace{0.5cm}}}
        \toprule[2.0pt]
        \toprule[1.0pt]
        Parameter & Prior
        \\
        \midrule[1.0pt]
        ${M}_c$ & $\mathcal{U}\left[0.95{M}_c^{\rm inj}, 1.05{M}_c^{\rm inj}\right]$ \\
        $\eta$ & ${\cal U}(0.05,0.25)$ \\
        $\chi_1,\chi_2$ & ${\cal U}(-0.2,0.2)$ \\
        $d_L$ & uniform in source-frame volume \\
        $\Phi_c$ & ${\cal U}(-\pi,\pi)$\\
        $t_c$ & ${\cal U}(t_c^{\rm inj}-0.1\,{\rm s}, t_c^{\rm inj}+0.1\,{\rm s})$ \\
        $\tilde{\Lambda}$ & ${\cal U}(0,10^5)$ \\
        $\delta\tilde{\Lambda}$ & ${\cal U}(-5\times10^4,5\times10^4)$ \\
        $\lambda_f$ & ${\cal U}(0,1)$ \\
        \bottomrule[1.0pt]
        \bottomrule[2.0pt]
    \end{tabular}
    \caption{Priors adopted in the parameter estimation analyses.}
\label{tab:priors}
\end{table}

Since $\tilde{\Lambda}$ and $\delta\tilde{\Lambda}$ are sampled directly, not every point in the prior volume corresponds to physical component tidal deformabilities. At each likelihood evaluation we therefore reconstruct $\Lambda_2^{(1)}$ and $\Lambda_2^{(2)}$ and impose
\begin{equation}
\Lambda_2^{(1)}>0,
\qquad
\Lambda_2^{(2)}>0 .
\end{equation}
Points violating these conditions are excluded by assigning $\ln\mathcal{L}=-\infty$.

Unless stated otherwise, the nested-sampling runs use $N_{\rm live}=1000$ with the \texttt{acceptance-walk} sampling scheme.

\section{Representative posterior distributions}
\label{app:representative_corners}

\begin{figure*}[p]
    \centering
    \includegraphics[width=\textwidth]{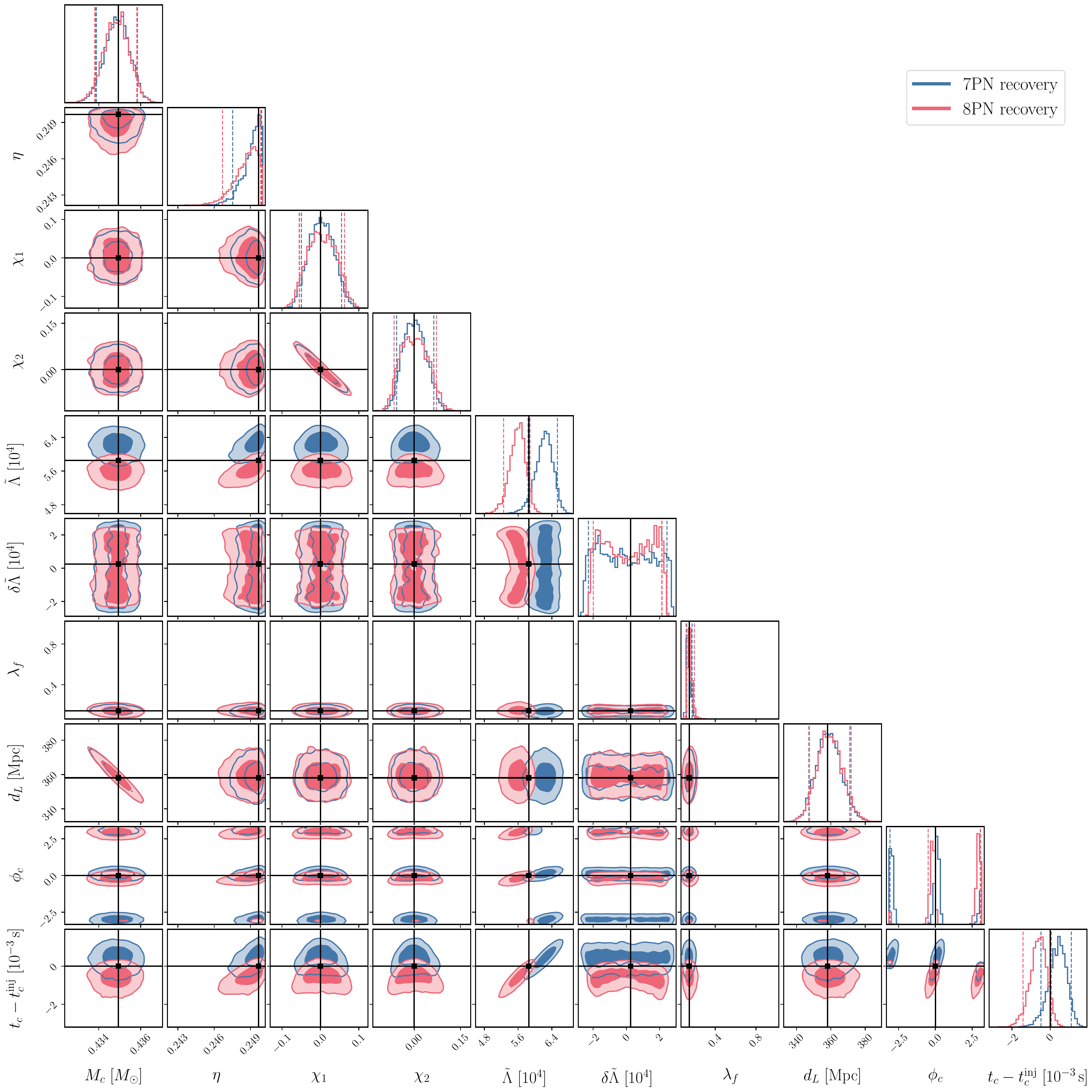}
    \caption{
    Posterior distributions of the sampled parameters for a binary with total mass $M_{\rm tot}=1\,M_\odot$, mass ratio $q=0.93$, and detector $\text{SNR}=50$. The injected signal includes tidal corrections through 8PN order and is recovered using either the full 8PN waveform (red) or a waveform truncated at 7PN order (blue). The inner and outer contours enclose $50\%$ and $90\%$ of the posterior probability, respectively, while dashed lines in the one-dimensional panels indicate the central $90\%$ credible intervals. Black lines mark the injected values. Here, $M_c$ denotes the source-frame chirp mass, and the coalescence time is measured relative to its injected value. The sky position, inclination, and polarization angle are fixed to their injected values (see Tab.~\ref{tab:fixed_injection_parameters}) and are therefore not shown.
    }
    \label{fig:corner_full_M1_snr50}
\end{figure*}

\begin{figure*}[p]
    \centering
    \includegraphics[width=\textwidth]{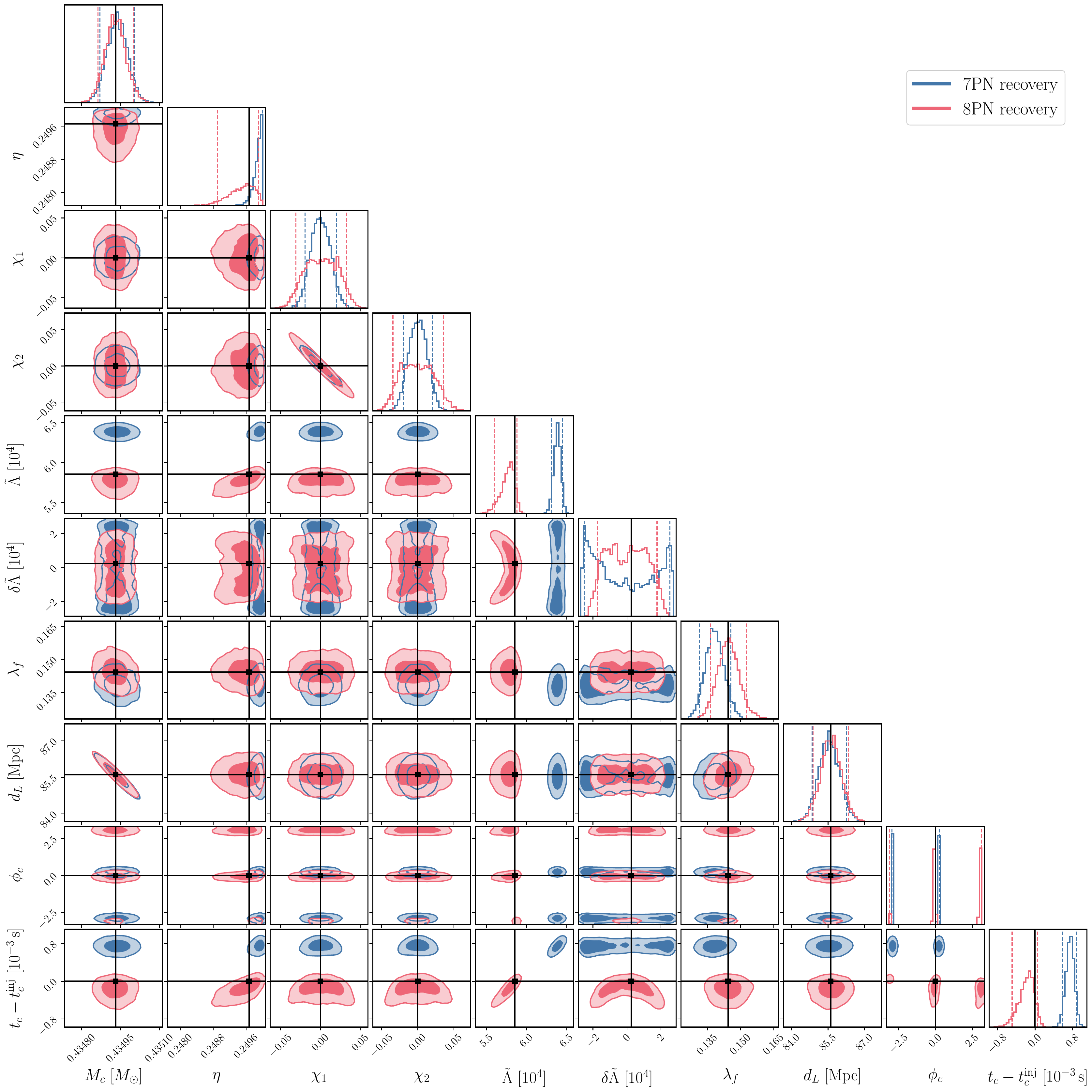}
    \caption{
    Same as Fig.~\ref{fig:corner_full_M1_snr50}, but for $M_{\rm tot}=1\,M_\odot$ and $\text{SNR}=200$.
    }
    \label{fig:corner_full_M1_snr200}
\end{figure*}

\begin{figure*}[p]
    \centering
    \includegraphics[width=\textwidth]{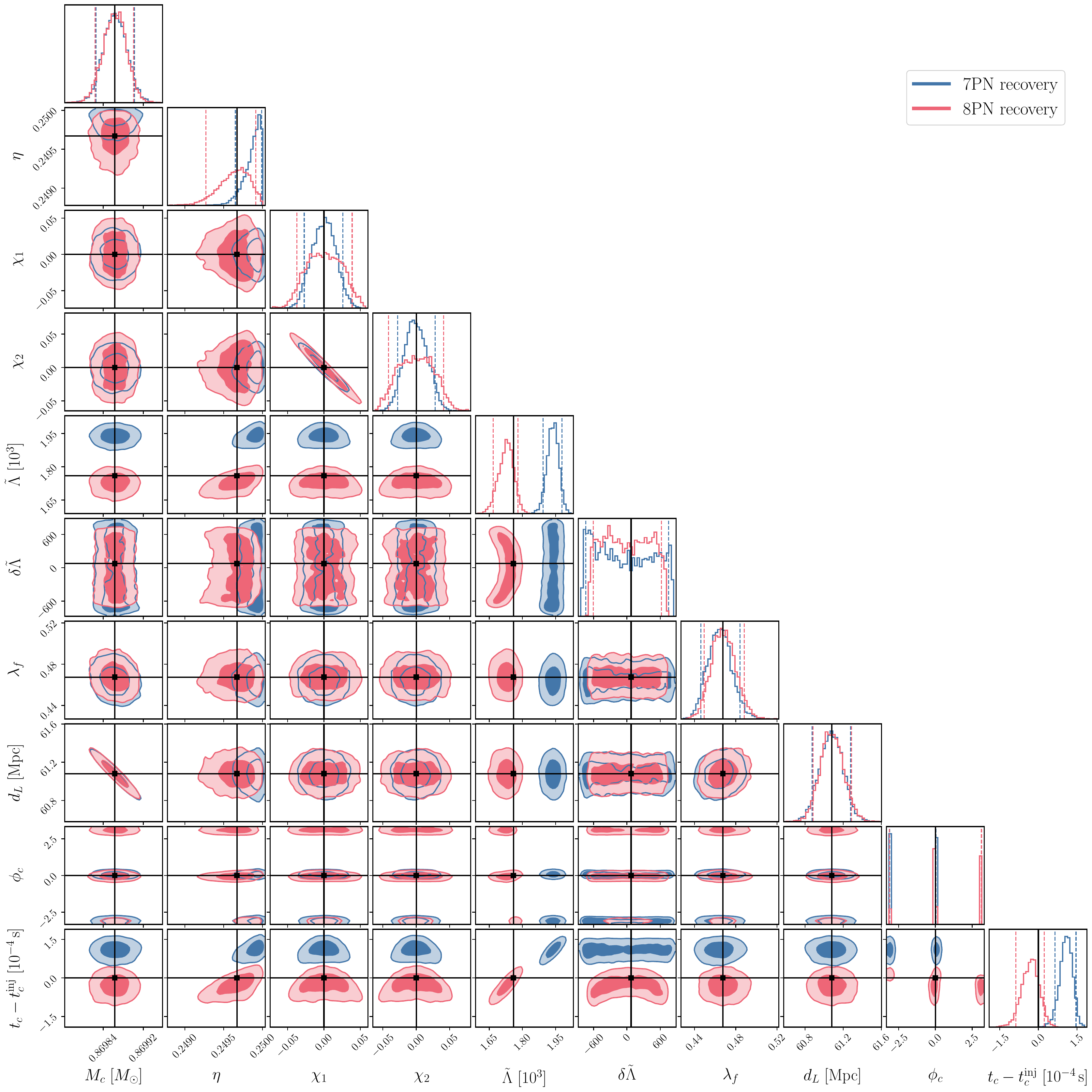}
    \caption{
    Same as Fig.~\ref{fig:corner_full_M1_snr50}, but for $M_{\rm tot}=2\,M_\odot$ and $\text{SNR}=500$.
    }
    \label{fig:corner_full_M2_snr500}
\end{figure*}

\begin{figure*}[p]
    \centering
    \includegraphics[width=\textwidth]{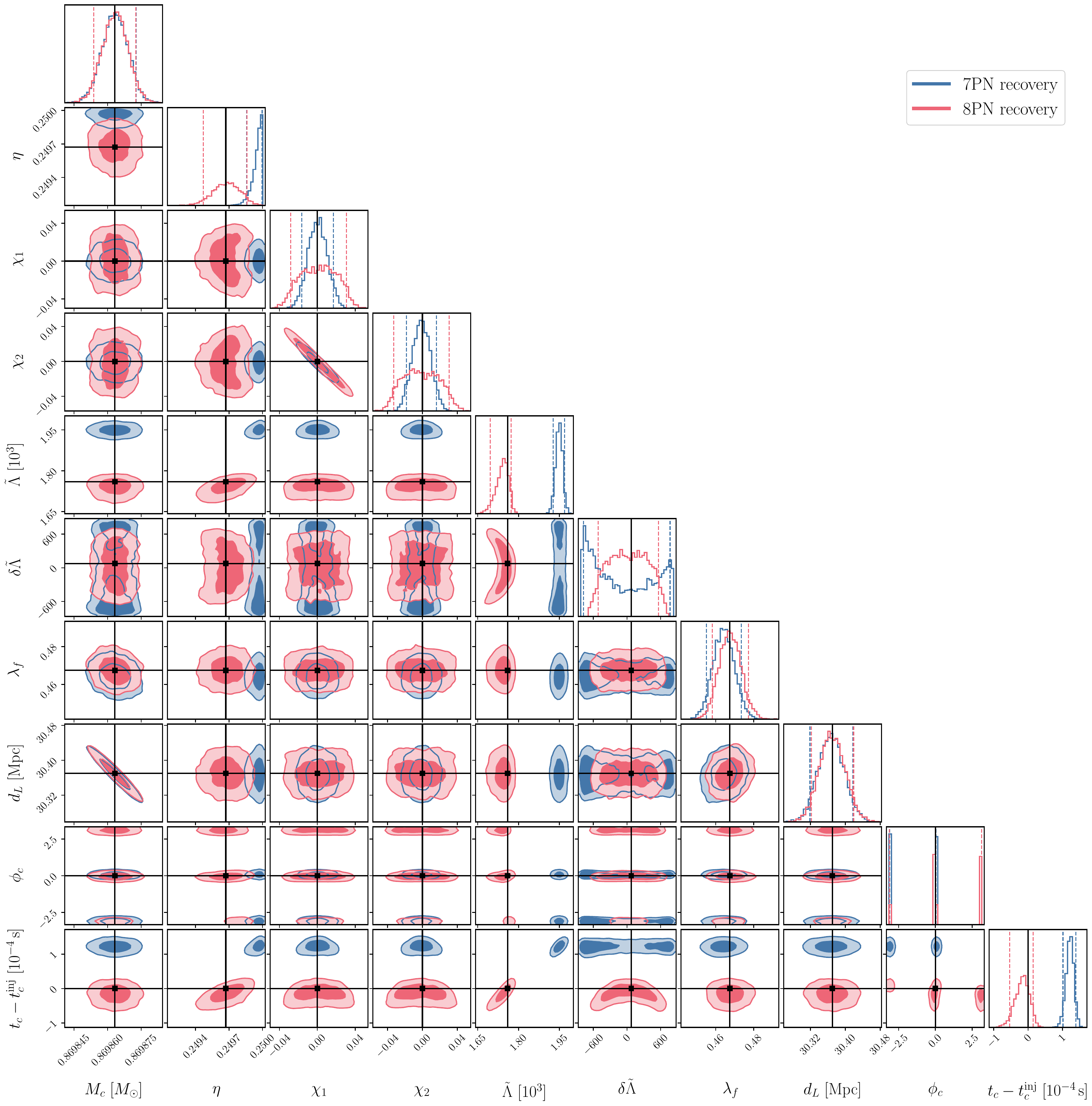}
    \caption{
    Same as Fig.~\ref{fig:corner_full_M1_snr50}, but for $M_{\rm tot}=2\,M_\odot$ and $\text{SNR}=1000$.
    }
    \label{fig:corner_full_M2_snr1000}
\end{figure*}

For completeness, Figs.~\ref{fig:corner_full_M1_snr50}--\ref{fig:corner_full_M2_snr1000} show the posterior distributions of all sampled parameters for representative injections generated with the full model and recovered with either the full model or the waveform truncated at 7PN. We consider nearly equal-mass binaries ($q=0.93$) with $M_{\rm tot}=1\,M_\odot$ at $\text{SNR}=50$ and $200$ (Figs.~\ref{fig:corner_full_M1_snr50} and~\ref{fig:corner_full_M1_snr200}), and with $M_{\rm tot}=2\,M_\odot$ at $\text{SNR}=500$ and $1000$ (Figs.~\ref{fig:corner_full_M2_snr500} and~\ref{fig:corner_full_M2_snr1000}). These cases range from substantial to decisive evidence for the 8PN terms (see Fig.~\ref{fig:bayes-factor-8PN-vs-7PN}). Figure~\ref{fig:corner_full_M1_snr200} shows the full set of sampled parameters for the injection of Fig.~\ref{fig:corner_qur_M2p8}.

In all cases, the two recoveries differ mainly in $\tilde{\Lambda}$ and $\eta$, while the posteriors of the remaining parameters are nearly identical.

\clearpage
\bibliography{biblio}

\end{document}